\documentclass[sn-nature]{sn-jnl}

\usepackage{graphicx}%
\usepackage{multirow}%
\usepackage{amsmath,amssymb,amsfonts}%
\usepackage{amsthm}%
\usepackage{mathrsfs}%
\usepackage[title]{appendix}%
\usepackage{xcolor}%
\usepackage{textcomp}%
\usepackage{manyfoot}%
\usepackage{booktabs}%
\usepackage{algorithm}%
\usepackage{algorithmicx}%
\usepackage{algpseudocode}%
\usepackage{listings}%
\usepackage{subcaption}
\usepackage{siunitx}
\usepackage[normalem]{ulem}

\newcommand{\blue}{\textcolor{blue}}

\newcommand{\dg}{$^{\circ}$}
\newcommand{\micron}{$\mu$m}
\newcommand{\um}{$\mu$m}

\newcommand\redout{\bgroup\markoverwith
{\textcolor{red}{\rule[.5ex]{2pt}{0.8pt}}}\ULon}

\newcommand{\nitrogen}{N$_2$}
\newcommand{\coo}{CO$_2$}
\newcommand{\ethane}{C$_2$H$_6$}
\newcommand{\dmethane}{CH$_3$D}
\newcommand{\methane}{CH$_4$}

\newcommand{\acet}{C$_2$H$_2$}
\newcommand{\methyl}{CH$_3$}

\def\ch#1{CH$_{#1}$}  
\def\cms{cm$^3$\,s$^{-1}$}
\def\e#1{$\times$10$^{#1}$}
\def\n2{N$_2$}
\def\nh3{NH$_3$}
\def\ni#1{$\nu_{#1}$} 
\def\v#1{$v_{#1}$} 
\def\k#1{$k_{#1}$}

\newcommand{\jqsrt}{J. Quant. Spectrosc. Radiat. Transfer} 
\newcommand{\icarus}{Icarus} 

\begin{document}

\title[The Atmosphere of Titan in Late Northern Summer from JWST and Keck Observations]{The Atmosphere of Titan in Late Northern Summer from JWST and Keck Observations}



\author*[1]{\fnm{Conor A.} \sur{Nixon}}\email{conor.a.nixon@nasa.gov}

\author[2]{\fnm{Bruno} \sur{B\'ezard}}\email{bruno.bezard@obspm.fr}
\equalcont{These authors contributed equally to this work.}

\author[3]{\fnm{Thomas} \sur{Cornet}}\email{thomas.cornet@esa.int}
\equalcont{These authors contributed equally to this work.}

\author[4]{\fnm{Brandon Park} \sur{Coy}} \email{bpcoy@uchicago.edu}
\equalcont{These authors contributed equally to this work.}

\author[5]{\fnm{Imke}\sur{de Pater}}\email{imke@berkeley.edu}
\equalcont{These authors contributed equally to this work.}

\author[6,7]{\fnm{Maël} \sur{Es-Sayeh}}\email{mael.es-sayeh@meteo.fr}
\equalcont{These authors contributed equally to this work.}

\author[8]{\fnm{Heidi B.} \sur{Hammel}} \email{hbhammel@aura-astronomy.org}
\equalcont{These authors contributed equally to this work.}

\author[2]{\fnm{Emmanuel} \sur{Lellouch}}\email{emmanuel.lellouch@obspm.fr}
\equalcont{These authors contributed equally to this work.}

\author[9]{\fnm{Nicholas A.} \sur{Lombardo}}\email{nicholas.lombardo@yale.edu}
\equalcont{These authors contributed equally to this work.}

\author[10]{\fnm{Manuel} \sur{L\'opez-Puertas}} \email{puertas@iaa.es}
\equalcont{These authors contributed equally to this work.}

\author[9]{\fnm{Juan M.}\sur{Lora}}\email{juan.lora@yale.edu}
\equalcont{These authors contributed equally to this work.}

\author[11]{\fnm{Pascal} \sur{Rannou}}\email{pascal.rannou@univ-reims.fr}
\equalcont{These authors contributed equally to this work.}

\author[6]{\fnm{Sébastien} \sur{Rodriguez}}\email{rodriguez@ipgp.fr}
\equalcont{These authors contributed equally to this work.}

\author[12]{\fnm{Nicholas A.} \sur{Teanby}} \email{n.teanby@bristol.ac.uk}
\equalcont{These authors contributed equally to this work.}

\author[13]{\fnm{Elizabeth P.} \sur{Turtle}} \email{elizabeth.turtle@jhuapl.edu}
\equalcont{These authors contributed equally to this work.}


\author[14]{\fnm{Richard K.} \sur{Achterberg}} \email{richard.k.achterberg@nasa.gov}

\author[15]{\fnm{Carlos} \sur{Alvarez}} \email{calvarez@keck.hawaii.edu}

\author[16]{\fnm{Ashley G.} \sur{Davies}} \email{ashley.g.davies@jpl.nasa.gov}

\author[17]{\fnm{Katherine} \sur{de~Kleer}} 
\email{dekleer@caltech.edu}

\author[15]{\fnm{Greg} \sur{Doppmann}} \email{gdoppmann@keck.hawaii.edu}

\author[18]{\fnm{Leigh N.} \sur{Fletcher}} \email{leigh.fletcher@leicester.ac.uk}

\author[19]{\fnm{Alexander~G.} \sur{Hayes}} \email{hayes@astro.cornell.edu}

\author[20]{\fnm{Bryan J.} \sur{Holler}} 
\email{bholler@stsci.edu}

\author[21]{\fnm{Patrick G. J.} \sur{Irwin}} \email{patrick.irwin@physics.ox.ac.uk}

\author[15]{\fnm{Carolyn} \sur{Jordan}} 
\email{cjordan@keck.hawaii.edu}

\author[18]{\fnm{Oliver R. T.} \sur{King}} \email{ortk2@leicester.ac.uk}

\author[22]{\fnm{Nicholas W.} \sur{Kutsop}} \email{nkutsop@uidaho.edu}

\author[18]{\fnm{Theresa} \sur{Marlin}} \email{henrik.melin@leicester.ac.uk}

\author[18]{\fnm{Henrik} \sur{Melin}} \email{tmarlin@caltech.edu}

\author[23]{\fnm{Stefanie N.} \sur{Milam}} \email{stefanie.milam@nasa.gov}

\author[20]{\fnm{Edward M.} \sur{Molter}} \email{ned.molter@gmail.com}

\author[24]{\fnm{Luke} \sur{Moore}} \email{moore@bu.edu}

\author[10]{\fnm{Yaniss} \sur{Nyffenegger-P\'er\'e}} \email{yanissnp@iaa.es}

\author[25]{\fnm{James} \sur{O'Donoghue}} \email{james.odonoghue@reading.ac.uk}

\author[15]{\fnm{John} \sur{O'Meara}} \email{jomeara@keck.hawaii.edu}

\author[26]{\fnm{Scot C. R.} \sur{Rafkin}} \email{scot.rafkin@swri.org}

\author[18]{\fnm{Michael T.} \sur{Roman}} \email{m.t.roman@leicester.ac.uk}

\author[15]{\fnm{Arina} \sur{Rostopchina}} \email{arostopchina@keck.hawaii.edu}

\author[25]{\fnm{Naomi} \sur{Rowe-Gurney}} \email{}

\author[23]{\fnm{Carl} \sur{Schmidt}} \email{schmidttc@bu.edu}

\author[27]{\fnm{Judy} \sur{Schmidt}} \email{geckzilla@gmail.com}

\author[28]{\fnm{Christophe} \sur{Sotin}} \email{christophe.sotin@uiv-nantes.fr}

\author[29]{\fnm{Tom S.} \sur{Stallard}} \email{tom.stallard@northumbria.ac.uk}

\author[20]{\fnm{John A.} \sur{Stansberry}} \email{jstans@stsci.edu}

\author[16]{\fnm{Robert A.} \sur{West}} \email{rwest15@gmail.com}




\affil*[1]{\orgdiv{Planetary Systems Laboratory}, \orgname{NASA Goddard Space Flight Center}, \orgaddress{\street{8800 Greenbelt Road}, \city{Greenbelt}, \postcode{20771}, \state{MD}, \country{USA}}}

\affil[2]{\orgdiv{LIRA}, \orgname{Observatoire de Paris, Universit\'e PSL, Sorbonne Universit\'e, Universit\'e Paris Cit\'e, CY Cergy Paris Universit\'e, CNRS}, \orgaddress{\street{Place Jules Janssen}, \city{Meudon}, \postcode{92195}, \country{France}}}

\affil[3]{\orgdiv{European Space Agency (ESA)}, \orgname{European Space Astronomy Centre (ESAC)}, \orgaddress{\street{Camino Bajo del Castillo s/n, Urb. Villafranca del Castillo}, \city{Villanueva de la Cañada}, \postcode{28692}, \country{Spain}}}

\affil[4]{\orgdiv{Department of Geophysical Sciences}, \orgname{University of Chicago}, \orgaddress{5734 S. Ellis Ave}, \city{Chicago}, \postcode{60615}, \state{IL}, \country{USA}}

\affil[5]{\orgdiv{Department of Astronomy and Department of Earth and Planetary Science}, \orgname{University of California Berkeley}, \orgaddress{501 Campbell Hall}, \city{Berkeley}, \postcode{94720}, \country{USA}}

\affil[6]{
\orgname{Université Paris Cité, Institut de Physique du Globe de Paris (IPGP), CNRS}, \orgaddress{\city{Paris}, \postcode{75005},  \country{France}}}

\affil[7]{
\orgname{CNRM, Météo-France, CNRS}, \orgaddress{\city{Toulouse}, \postcode{31100}, \country{France}}}

\affil[8]{
\orgname{Association of Universities for Research in Astronomy}, \orgaddress{\street{1331 Pennsylvania Ave NW}, \city{Washington}, \postcode{DC 20004}, 
\country{USA}}}

\affil[9]{\orgdiv{Department of Earth and Planetary Sciences}, \orgname{Yale University}, \orgaddress{\street{210 Whitney Avenue}, \city{New Haven}, \postcode{CT 06511}, 
\country{USA}}}

\affil[10]{\orgdiv{Instituto de Astrofísica de Andalucía}, \orgname{CSIC}, \orgaddress{Glorieta de la Astronomía, s/n}, \city{Granada}, \postcode{E-18008}, \country{Spain}}

\affil[11]{\orgdiv{GSMA, UMR CNRS 7331}, \orgname{Université de Reims Champagne-Ardenne}, \orgaddress{\street{Campus Sciences Exactes et Naturelles}, \city{REIMS}, \postcode{51687}, \state{}, \country{France}}}

\affil[12]{\orgdiv{School of Earth Sciences}, \orgname{University of Bristol}, \orgaddress{Queens Road}, \city{Bristol}, \postcode{BS8 1RJ}, \country{UK}}

\affil[13]{\orgdiv{Space Exploration Sector}, \orgname{Johns Hopkins Applied Physics Laboratory}, \orgaddress{\street{11100 Johns Hopkins Road}, \city{Laurel}, \postcode{20723}, \state{MD}, \country{USA}}}


\affil[14]{\orgdiv{Department of Astronomy/CRESST}, \orgname{University of Maryland}, \orgaddress{\street{4296 Stadium Dr.}, 
\city{College Park}, \postcode{20742}, \state{MD}, \country{USA}}}

\affil[15]{
\orgname{W.M. Keck Observatory}, \orgaddress{\street{65-1120 Mamalahoa Hwy}, \city{Kamuela}, \postcode{HI 96743}, 
\country{USA}}}

\affil[16]{
\orgname{Jet Propulsion Laboratory}, \orgaddress{\street{4800 Oak Grove Drive}, \city{Pasadena}, \postcode{CA 91011}, 
\country{USA}}}

\affil[17]{\orgdiv{Division of Geological and Planetary Sciences}, \orgname{Caltech}, \orgaddress{1200 E. California Blvd}, \city{Pasadena}, \postcode{CA 91125}, 
\country{USA}}

\affil[18]{\orgdiv{School of Physics and Astronomy}, \orgname{University of Leicester}, \orgaddress{\street{University Road}, \city{Leicester}, \postcode{LE1 7RH}, 
\country{UK}}}

\affil[19]{\orgdiv{Department of Astronomy}, \orgname{Cornell University}, \orgaddress{\street{122 Sciences Drive}, \city{Ithaca}, \postcode{NY 14850}, 
\country{USA}}}

\affil[20]{
\orgname{Space Telescope Science Institute}, \orgaddress{\street{3700 San Martin Drive}, \city{Baltimore}, \postcode{MD 21218}, 
\country{USA}}}

\affil[21]{\orgdiv{Atmospheric, Oceanic and Planetary Physics}, \orgname{University of Oxford}, \orgaddress{\street{Parks Road}, \city{Oxford}, \postcode{OX1 3PU}, \state{England}, \country{UK}}}

\affil[22]{\orgdiv{Department of Physics}, \orgname{University of Idaho}, \orgaddress{\street{875 Perimeter Drive}, \city{Moscow}, \postcode{ID 83844}, 
\country{USA}}}

\affil[23]{\orgdiv{Astrochemistry Laboratory}, \orgname{NASA Goddard Space Flight Center}, \orgaddress{\street{8800 Greenbelt Road}, \city{Greenbelt}, \postcode{20771}, \state{MD}, \country{USA}}}

\affil[24]{\orgdiv{Dept. of Astronomy}, 
\orgname{Boston University}, \orgaddress{\street{725 Commonwealth Ave}, \city{Boston}, \postcode{MA 02215}, 
\country{USA}}}

\affil[25]{\orgdiv{Department of Meteorology}, \orgname{University of Reading}, \orgaddress{\street{Whiteknights Road}, \city{Reading}, \postcode{RG6 6ET}, 
\country{UK}}}

\affil[26]{
\orgname{Southwest Research Institute}, \orgaddress{\street{1050 Walnut St \#300}, \city{Boulder}, \postcode{CO 80302}, 
\country{USA}}}

\affil[27]{
\orgname{Independent Astronomer}, 
\orgaddress{
\state{CA}, \country{USA}}}

\affil[28]{\orgdiv{Laboratoire de Planétologie et Géodynamique, CNRS UMR 6112}, \orgname{Nantes University}, \orgaddress{ \city{Nantes}, \postcode{44322},  \country{France}}}

\affil[29]{\orgdiv{Northumbria University City Campus}, 
 \orgaddress{ \city{Newcastle upon Tyne}, \postcode{NE1 8ST},  \country{UK}}}



\abstract{


Saturn’s moon Titan undergoes a long annual cycle of 29.45 Earth years. Titan's northern winter and spring were investigated in detail by the Cassini-Huygens spacecraft (2004-2017), but the northern summer season remains sparsely studied. Here we present new observations from the James Webb Space Telescope (JWST) and Keck II telescope made in 2022 and 2023 during Titan’s late northern summer. With JWST’s Mid-Infrared Instrument we spectroscopically detected the methyl radical, the primary product of methane breakup and key to the formation of ethane and heavier molecules. With JWST's  Near Infrared Spectrograph we detected several non-LTE CO and CO2 emission bands that allow measurement of these species over a wide altitude range. Lastly, with JWST's Near Infrared Camera and Keck II we imaged northern hemisphere tropospheric clouds evolving in altitude, providing new insights and constraints on seasonal convection patterns. These observations pave the way for new observations and modeling of Titan's climate and meteorology as it progresses through northern Fall equinox, when its atmosphere is expected to show dramatic seasonal changes.
 
}




\maketitle

\section{Main Text}\label{sect:intro}


Saturn's moon Titan is the only moon in the solar system with a dense atmosphere, composed largely of nitrogen (94.4--98.5\%) and methane (5.5--1.4\%) \cite{niemann2010composition}. Due to the cold troposphere and surface (70--93~K) \cite{fulchignoni2005situ} close to the triple point of methane, methane is a condensable substance in Titan's lower atmosphere. It therefore plays a similar meteorological role to water on Earth, evaporating from the surface and reaching the middle troposphere, where methane clouds form and rainfall occurs in changing seasonal patterns \cite{turtle2011seasonal, Rodriguez2009titan,rodriguez2011titan,turtle2018titan}. 

In the middle and upper atmosphere, methane along with nitrogen is processed by photochemistry \cite{yung1984,horst2017}, leading to the formation of heavier organic molecules (e.g. C$_2$H$_2$, C$_2$H$_6$, HCN) along with larger PAHs (polycyclic aromatic hydrocarbons) \cite{lopezpuertas2013} and eventually solid organic haze particles. These particles are thought to be the building blocks of stratospheric aerosols \cite{lavvas2013} as they drift downwards and grow in size to form a global haze layer in the stratosphere, which lends Titan its golden hue at visible wavelengths. In the stratosphere, seasonal changes in composition and haze distribution \cite{lorenz2004seasonal,West2018,Mathe2020,seignovert2021} indicate the seasonal reversal of Titan's global meridional circulation every 14.75 Earth years \cite{lebonnois2001seasonal,hourdin2004titan,lebonnois2012,lombardo2023}.
 
Since the end of the Cassini-Huygens mission in September 2017 our ability to monitor Titan's changing atmosphere has been curtailed, since we have lacked the ability to view Titan unimpeded by the Earth's atmosphere across all infrared wavelengths. Nevertheless, using ground-based adaptive optics (AO) imaging from observatories such as Keck on Maunakea, cloud monitoring was possible, although only in some near-infrared (NIR) spectral windows. With the commissioning of the James Webb Space Telescope (JWST) in 2022
it became possible to view Titan once more at all wavelengths from $\sim$0.7 to 28.0~\micron\ clearly, from space \cite{clampin2008james}.

In this article we report on the first results of JWST imaging and spectroscopy of Titan that occurred in November 2022 and July 2023, using three of its four  instruments.
We also report on contemporaneous ground-based observations with the Keck II NIRC2 camera that provided even higher spatial resolution imaging over a wider time range, but in a more limited set of filters. The observations and data reduction are described in detail in the Supplementary Online Material A and B.
Together, these results provide a new, integrated look at the composition and meteorology of Titan's atmosphere 
in 2022-2023 (L$_s$=150-158$^{\circ}$) from the upper atmosphere to the surface, at a season that was poorly documented by previous observations (\blue{Extended Data Figure 1}).


\section{Results}\label{results}


\subsection{Detection of the \methyl\ radical on Titan}

\begin{figure}[hb]
    \centering
    \includegraphics[width=1.0\linewidth]{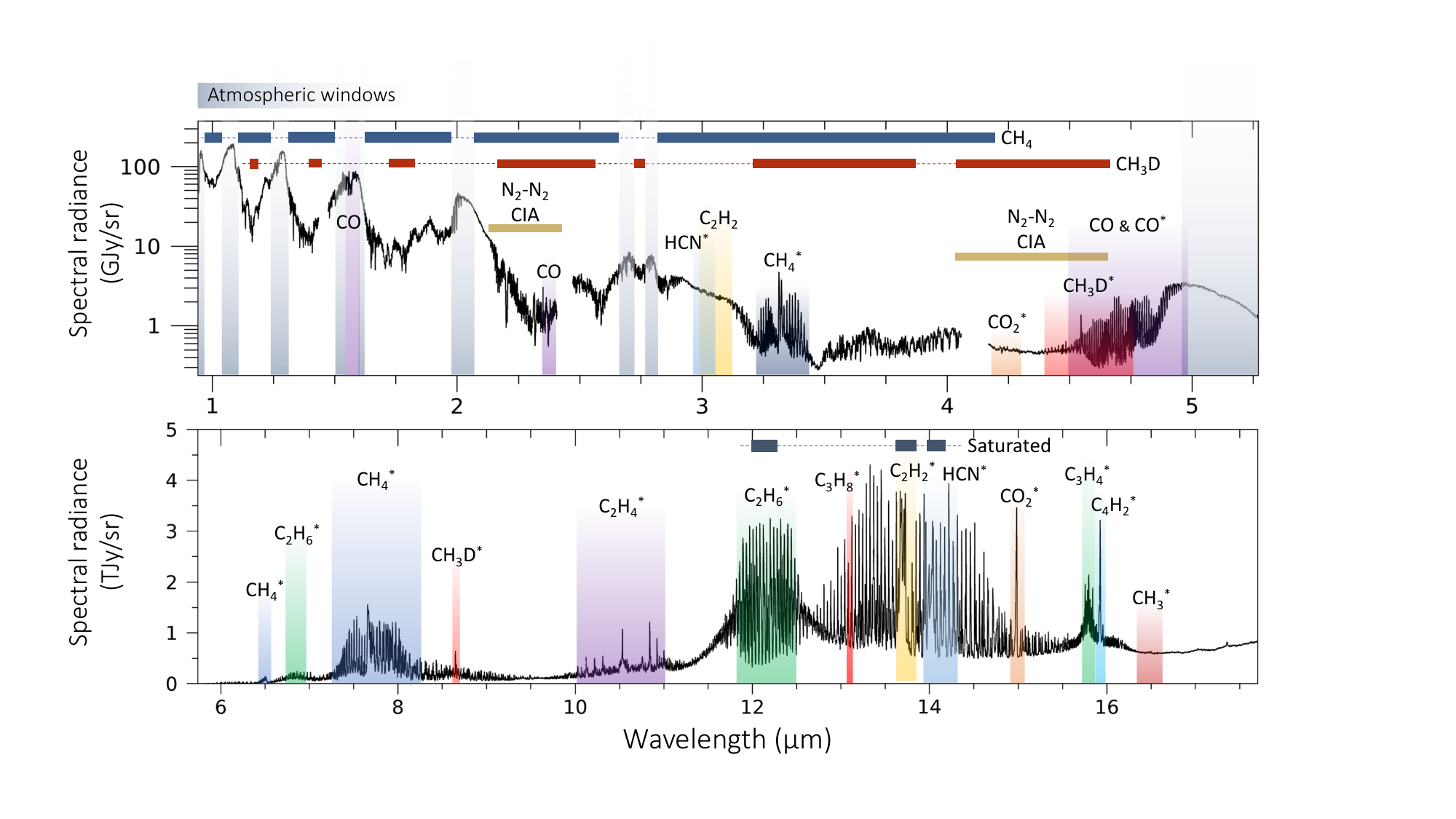}
    \caption{{\bf Disk-averaged JWST spectra of Titan}. Top: NIRSpec.  Bottom: MIRI (Channels 1B - 3C). Main absorption and emission bands (the latter labeled with a *) are shown, as well as spectral transmission windows where Titan's surface can be seen (grey shaded regions).}
    \label{fig:combined-spectrum}
\end{figure}

Observations of Titan with JWST's Mid Infrared Instrument (MIRI) were conducted on July 11th 2023 in Medium-Resolution Spectrometer (MRS) mode ($R \sim 10^3$) (see Supplementary \blue{Table 1})
for observational parameters). The lower panel of \blue{Figure~\ref{fig:combined-spectrum}} shows the disk-averaged spectrum, covering most of the mid-infrared spectrum available to the MRS instrument, from Channel 1B to 3C (5.7 to 18.0 microns). 
Channel 1A and Channel 4A-C are not shown due to ongoing calibration considerations and/or low SNR (signal-to-noise ratio), and will be addressed in a later paper. 

The exceptional signal-to-noise ratio across much of the central MIRI/MRS spectrum has permitted JWST to make the first ever spectroscopic detection of the CH$_3$ radical in Titan's atmosphere. The methyl radical is one of the most important photochemical dissociation products of methane and much of the hydrocarbon chemistry in Titan's atmosphere cascades from this species \cite{19vuietal}. It is also the main source of CO on Titan, stemming from its upper atmosphere ($\sim$1000 km) reaction with either atomic oxygen \citep{Horst2008, Lara2014} or OH \citep{19vuietal}. In addition, the CH$_3$ radical leads to the creation of many of Titan's higher order photochemical products (C$_2$, C$_3$, C$_4$, $\ldots$) - and in particular ethane (C$_2$H$_6$) which is a significant component of polar clouds \citep{griffith2006evidence}, as well as the lakes and seas \citep{brown2008identification, cordier2009estimate, mastrogiuseppe2016radar, mastrogiuseppe2019deep}.

Previously, the only constraint on the concentration of the methyl radical was the detection of the related ion species CH$_3^+$ in Titan's ionosphere above $\sim$1000~km with Cassini/INMS \cite[e.g.][]{12manetal}, but its signature is combined with the NH$^+$ ion that has the same mass, leading to a large uncertainty. 

\begin{figure}
    \centering
    \includegraphics[width=1.0\linewidth] {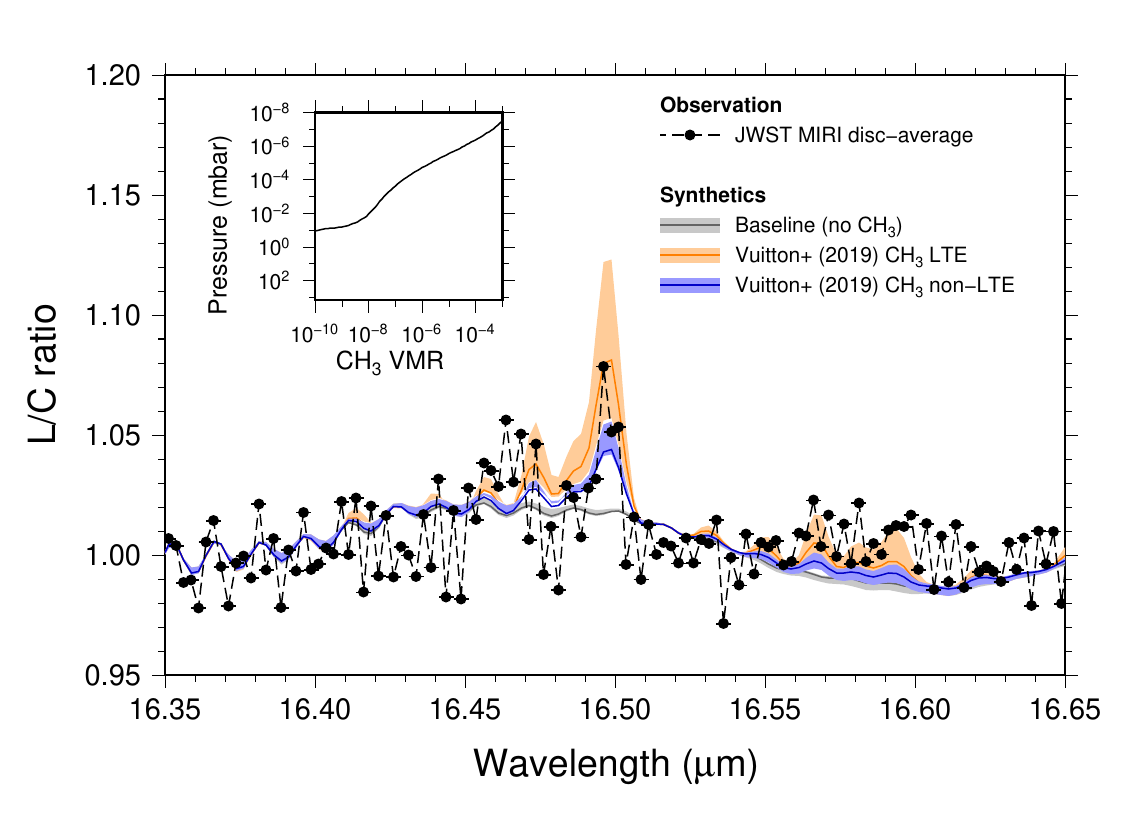}
    \caption{{\bf Detection of the methyl (CH$_3$) radical on Titan with MIRI/MRS.} Observed and modeled spectra have been converted to a line-to-continuum (L/C) ratio by division with a second order polynomial fitted to the continuum regions.  
    Synthetic spectra are based on a predicted CH$_3$ profile (inset) from a photochemical model \cite{19vuietal}. Forward modeling errors on the synthetic spectra result from the combined uncertainty in the temperature-pressure profile and the non-LTE collisional rate coefficient k$_1$ (see Sec.\,\ref{sec:mirimodel}). Observational errors on the L/C ratio are $<$0.001, so are negligible compared to the forward model error.  There is however significant residual rippling on the data after running the standard JWST pipeline (see Supplement B.1), which is estimated to be $\sim$0.0135 using the root mean square ripple amplitude of the 16.35--16.46~$\mu$m continuum region. The fit to the JWST spectrum is much improved when CH$_3$ emission is included. Normalised misfits ($\chi^2/n$) of the three modelled spectra are: 1.62 for the baseline; 1.40 for the LTE case; and 1.32 when non-LTE effects are included. This indicates good CH$_3$ abundance agreement between observations and photochemical model predictions.}
    \label{fig:miri-figure}
\end{figure}

In \blue{Fig.~\ref{fig:miri-figure}} we show the spectroscopic detection of CH$_3$, where a forward model spectrum (Methods \ref{sec:mirimodel}) based on a photochemical model prediction of the CH$_3$ profile \cite{19vuietal} is compared to the disk-averaged JWST MIRI/MRS spectrum. 
Uncertainties in the model originate from uncertainties in the atmospheric temperature and 
non-local thermodynamic equilibrium (non-LTE) effects at very low thermospheric pressures ($<10^{-3}$~mbar).

The CH$_3$ model spectra with LTE and non-LTE source functions provide comparable fits to the observations, with the LTE model slightly over-fitting and the non-LTE model slightly under-fitting (\blue{Fig.~\ref{fig:miri-figure}}). 
However, our non-LTE calculations show that above 500~km the number of CH$_3$ radicals in the excited upper state is much less than in the LTE case (see \blue{Methods \ref{sec:mirimodel}} and \blue{Extended Data Fig.~2}).
This suggests that emission from the thermosphere is minimal and observations are sensitive mainly to CH$_3$ abundance in the stratopause region, where abundances are predicted to be sub-part per billion. 
The photochemical model is fully consistent with observations, which provides further confidence in current photochemical schemes that are extrapolated from Cassini/INMS measurements at  $z>$1000~km ($<$10$^{-6}$~mbar) down to stratospheric levels ($<$300~km, $>$0.1~mbar).


\clearpage 
\subsection{Near infrared daytime emissions of CO, \coo\ and \dmethane }\label{nirspec}


On November 4$^{th}$ 2022, the Integral Field Unit (IFU) mode of JWST's Near Infrared Spectrometer (NIRSpec) (see Supplement A.2) 
allowed us to observe Titan with the highest spectral resolution (R$\sim$2000-3500) \citep{Jakobsen2022} yet obtained across the full 1--5~\micron\ range, free from telluric absorption, and with very high SNR - see \blue{Fig.~\ref{fig:combined-spectrum}} (top panel).

This region is dominated by solar radiation absorbed by gases, scattered by the haze and, in regions of relative methane transparency, reflected by the surface. Besides methane ($^{12}$\methane, $^{13}$\methane\ and \dmethane), absorption features from CO, \acet\ and \ethane\ are present. 
In addition, emission from several gases in Titan's stratosphere (\methane, \dmethane, HCN, CO and \coo) is seen, which encodes unique information on their concentrations and Titan's temperature structure.

\begin{figure} [ht]
\hspace{-0.13\linewidth}
\includegraphics[width=0.87\linewidth, angle=270]{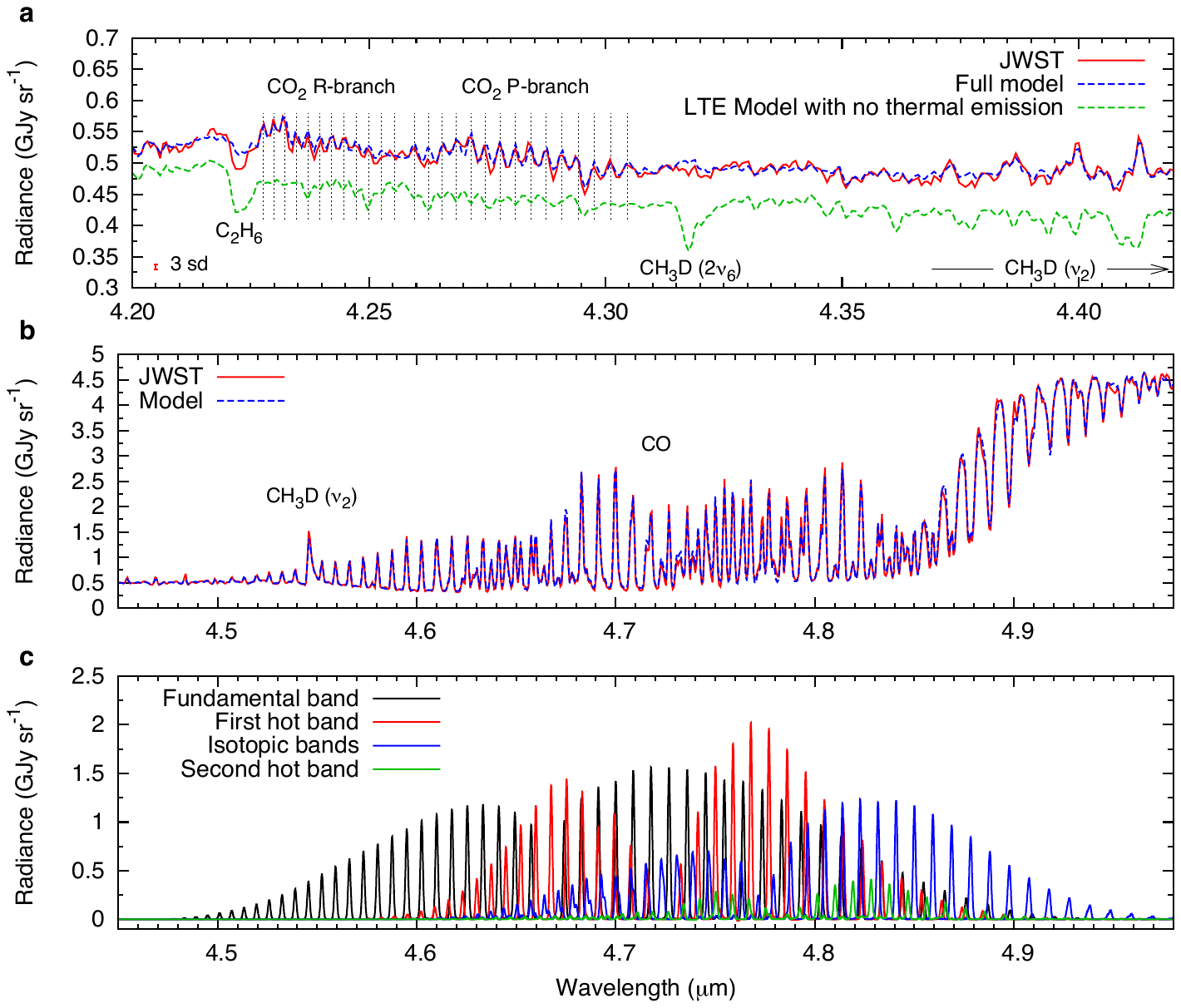} 
\caption{{\bf Bands of \coo, \dmethane, \ethane, and CO resolved by JWST NIRSpec.}
{\bf a}, an average of 36 NIRSpec spectra of Titan centered at 14{\dg}N in the 4.20--5.00 \micron\ spectral region (red line). 
The spectrum is compared with a synthetic spectrum including solar reflected radiation, thermal emission and non-LTE emission from \coo\ (blue dashed line), and with a synthetic spectrum without the thermal emission and the \coo\ non-LTE emission (green dashed line). \dmethane\ emission and absorption from the 2$\nu_6$ band near 4.32 \micron\ almost compensate each other. 
The vertical dotted lines indicate the location of the individual \coo\ lines. 
{\bf b}, The observed spectrum (red line) is compared with a synthetic spectrum including solar reflected radiation, thermal emission and non-LTE emission from CO only
(blue dashed line). {\bf c}, Contribution of the CO fundamental, first hot,  isotopic $^{13}$CO(1$\rightarrow$0) and C$^{18}$O(1$\rightarrow$0), and second hot bands to the CO emission.}
\label{fig:co_co2}
\end{figure}

\blue{Figure~\ref{fig:co_co2}a} shows the spectral features of Titan's outgoing radiation from 4.20 to 4.42\,\micron\ in an average of NIRSpec spectra around disk center (14\dg N). The spectrum is dominated by emission  of \coo\ lines of the strong 4.3\,\micron\ $\nu_3$ band. This is the first time that \coo\ emission from this band has been seen on Titan. A synthetic spectrum including only solar reflection and thermal emission does not produce enough radiance in the \coo\ lines. A temperature increase by some 5~K in the line forming region in LTE conditions (190-310~km) would be needed to reproduce the observed emission but such a model would overestimate the emission in the \dmethane\ bands mentioned below. 
Likewise, an increase of the CO$_2$ abundance could also explain this emission enhancement, but this is unlikely since it would need to be significantly larger than the CO$_2$ mole fraction profile retrieved from CIRS measurements \citep{Mathe2020}.
 However, a synthetic spectrum including solar pumped excitation of \coo(001), which produces a large enhancement of the \coo(001) vibrational temperature over the kinetic temperature above around 200\,km (see \blue{Extended Data Fig.~3}),
 reproduces the JWST spectrum very well (see Figure~\ref{fig:co_co2}a), hence suggesting that a large fraction of this \coo\ 4.3\,\micron\ emission occurs in non-LTE (see \blue{Methods \ref{sec:methods-nirspec}}).

The spectrum also exhibits an absorption from the weak $\nu_3$+$\nu_6$ band of \ethane\  near 4.22\,\micron\, which is partly filled in by thermal emission.  The \dmethane\ 2$\nu_6$ band near 4.32\,\micron\ is not visible as the absorption produced in the solar reflected component is compensated by thermal emission. \dmethane\ multiplets from the stronger $\nu_2$ band at 4.54\,\micron\ can be seen in emission beyond 4.37\,\micron. All these \dmethane\ features are  correctly reproduced by the LTE model.

\blue{Figure~\ref{fig:co_co2}b} shows the same daytime mean spectrum covering from 4.45 to 4.98\,\um. This region is dominated by \dmethane\ emission ($\nu_2$ band at 4.54\,\micron), which is correctly accounted for by thermal emission, by CO non-LTE 
emission across most of the range, and by CO absorption longwards of 4.85\,\um. 
This NIRSpec spectral selection has allowed us to measure the detailed structure of the CO 4.7\,\um\ non-LTE emission, including the fundamental, first and second hot bands of $^{12}$CO, and two isotopic bands with unparalleled quality (see \blue{Fig.~\ref{fig:co_co2}b--c}). In particular, it is the first time that the weak second hot band CO(3$\rightarrow$2) has been observed on Titan. Furthermore, the much higher spectral resolution of JWST spectra compared to those taken by Cassini/VIMS has enabled JWST to resolve the ro-vibrational lines
(see e. g., \blue{Fig.~1 in \cite{fabiano2017}}). 

The CO emission in the five identified bands occurs under non-LTE conditions. The CO emitting energy levels are excited by the absorption of solar radiation in the mid- and near- IR and emit at vibrational excitation temperatures considerably higher than the kinetic temperature (see \blue{Extended Data Fig.~3}).
We modelled the non-LTE radiances for the five CO bands (see \blue{Methods \ref{sec:methods-nirspec}}) and are able to reproduce the measured JWST spectrum very accurately, the residuals of the fit in the 4.46--4.86 $\mu$m interval being 0.068 GJy sr$^{-1}$. Panel {\bf b} of \blue{Fig.~\ref{fig:co_co2}} shows their total contribution together with the measured spectrum; their individual spectra are shown in Panel~{\bf c}. 

Besides the intrinsic value of knowledge of non-LTE processes, the discrimination of the different bands as well as the individual ro-vibrational lines of each band is important because 
the lines of these transitions have different opacities and hence allow us to retrieve the CO concentration profile over a wide range of altitudes, from about 50\,km up to about 700\,km (see \blue{Extended Data Fig.~4}).
This is significantly wider than that of Cassini/VIMS \cite{fabiano2017}. 
In the calculations shown in \blue{Fig.~\ref{fig:co_co2}} we used a constant-with-altitude CO volume mixing ratio (VMR) profile of 55 ppmv that matched all bands very well, including the absorption features beyond 4.85\,\um\ that probe the atmosphere down to the surface. In the stratosphere, our best-fit value of 55$\pm$5 ppmv is mostly based on the intensities of the hot band lines which are almost insensitive to temperature uncertainties. The 5 ppmv error bar derives from an estimated uncertainty in the flux calibration of 3\%.


Our derived CO concentration agrees very well (within the combined error bars) with the stratospheric abundance retrieved from ground-based millimeter and sub-millimeter  instruments (SMA, ALMA) \citep{Gurwell2004,Serigano2016}. 
In addition, several observations by CIRS have been reported to date, with CO abundances in the lower stratosphere (60-140\,km) of 45$\pm$15\,ppmv, 47$\pm$8\,ppmv and 55$\pm$6\,ppmv  \cite[][respectively]{Flasar2005,DeKok2007,Teanby2010}. 
Our observations are slightly larger than the first two and very close to the more recent of \cite{Teanby2010}. The SPIRE and PACS instruments on the Herschel space observatory also measured CO abundances in the lower stratosphere (60–170 km), obtaining values 
of 40$\pm$15 ppm \cite{Courtin2011} and 50$\pm$2 ppm \cite{Rengel2014}.
Our value is then also in very good agreement with these observations.
Further, it also agrees very well with the 
VMR of $\sim$60 ppmv more recently derived from VIMS limb spectra \citep{fabiano2017},
which extends up to $\sim$500\,km.
Thus, these observations suggest that the CO VMR is uniformly mixed with a constant value from the surface up to the lower thermosphere ($\sim$700\,km).
The vertical uniformity of CO, expected due to its long photochemical lifetime \citep[$\sim$500 MYr,][]{19vuietal}, is now verified over a greatly extended  vertical range, settling a long-standing debate on the distribution and longevity of CO in Titan's atmosphere. 



\subsection{Tropospheric Clouds}\label{clouds}

\begin{figure}[ht]\
\begin{centering}
\includegraphics[width=1.0\linewidth, angle=0]{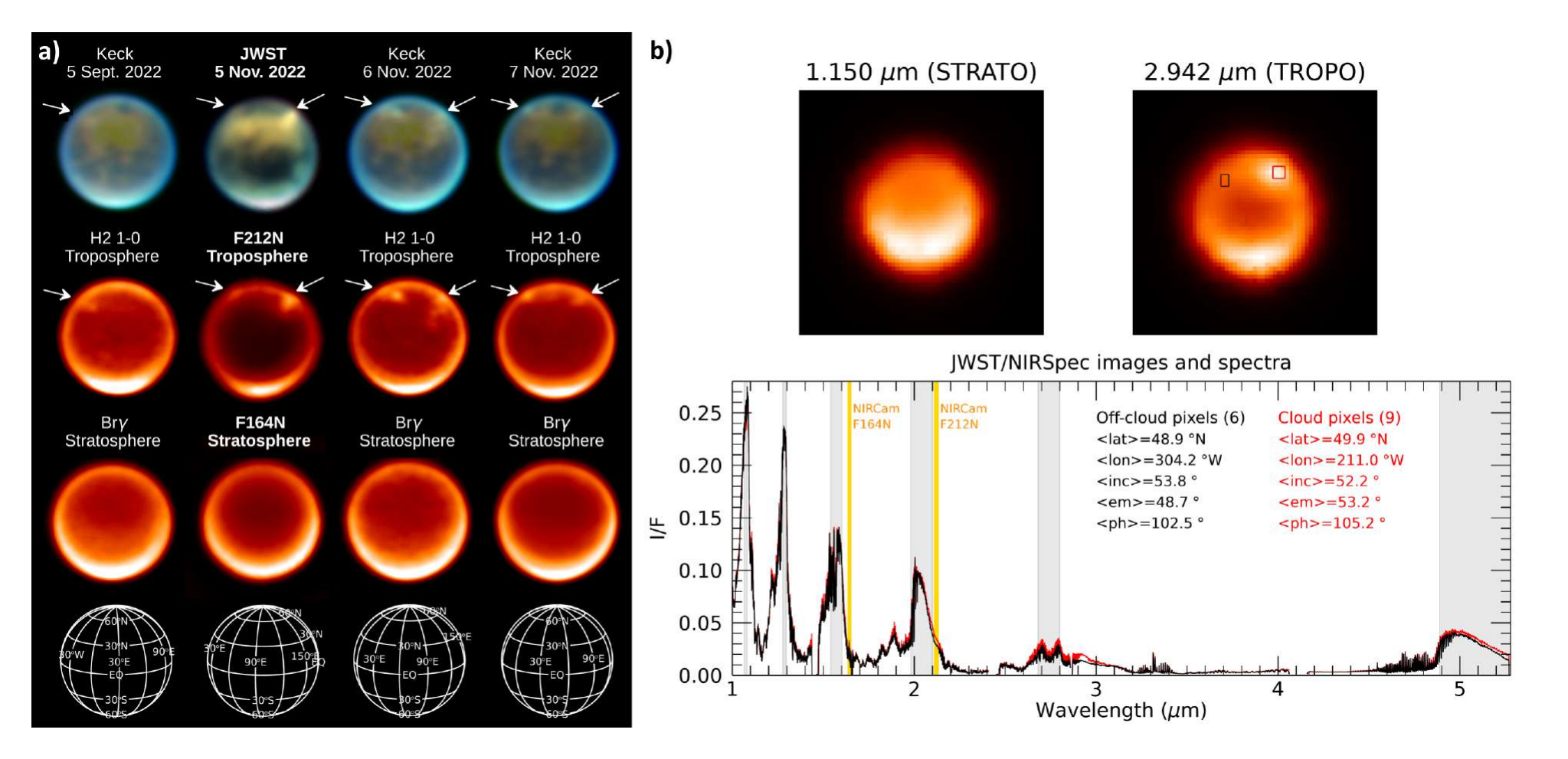}
\vspace{5mm}
\caption{{\bf Detection of clouds on Titan in 2022}
(a) Time series of JWST NIRCam and Keck II NIRC2 Titan cloud observations from September and November 2022. Row 1 shows RGBA color composite images.
JWST/NIRCam (5 Nov 2022): red - F200W; green - F164-F150W; blue - F140M; alpha - F210M.
Keck II/NIRC2 (other dates): red - He1b (2.056~\micron ); green - Kp (2.124~\micron ); blue - Br-$\gamma$ (2.169~\micron ); alpha - H2 $\nu =1{-}0$ (2.128~\micron ).
Rows 2 \& 3 show observations in single filters: row 2 shows troposphere sensing filters (F212N for NIRCam and H2 1-0 for NIRC2); row 3 shows filters sensitive primarily to the stratosphere (F164N for NIRCam and Br-$\gamma$ for NIRC2, with some contribution from the upper troposphere in the latter). White arrows indicate the location of clouds relative to surface features in rows 1 \& 2. Each individual filter image was pre-processed with an unsharp mask. Additionally, the F200W, F164N-F150W, and F210M images for NIRCam (He, Kp, H2 images for NIRC2) were empirically corrected for stratospheric haze contribution by subtracting a fraction of the F140M image (Br-$\gamma$ image for NIRC2). Row 4 shows the orientation of Titan at each epoch. See \blue{Fig.~\ref{fig:clouds_alts}} for the altitudes probed in the different filters. (b) JWST NIRSpec observation of Titan, for comparison to the NIRCam observation shown in (a). Images are shown at 1.150 \micron\ in a methane band, probing the stratosphere, and at 2.942 \micron , probing down to the troposphere. A cloud is clearly visible in the 2.942 \micron\ image, at the upper right, near Titan's limb (at the dusk side). At the bottom are shown two NIRSpec spectra between 1 and 5.28 \micron : (1) the spectrum in red corresponds to the average of 9 pixels located on the cloud, and (2) the spectrum in black corresponds to the average of 6 pixels off the cloud, at latitudes and observing geometries equivalent to the cloud spectrum. We show the `clear' atmospheric spectral windows as grey shaded regions. For comparison purposes, we also indicate the NIRCam tropospheric (F212N) and stratospheric (F164N) filters in orange, that are used to identify the clouds in (a).   
}
\label{fig:cloud_images1}
\end{centering}
\end{figure}

\begin{figure}[ht]\
\begin{centering}
\includegraphics[width=1.0\linewidth, angle=0]{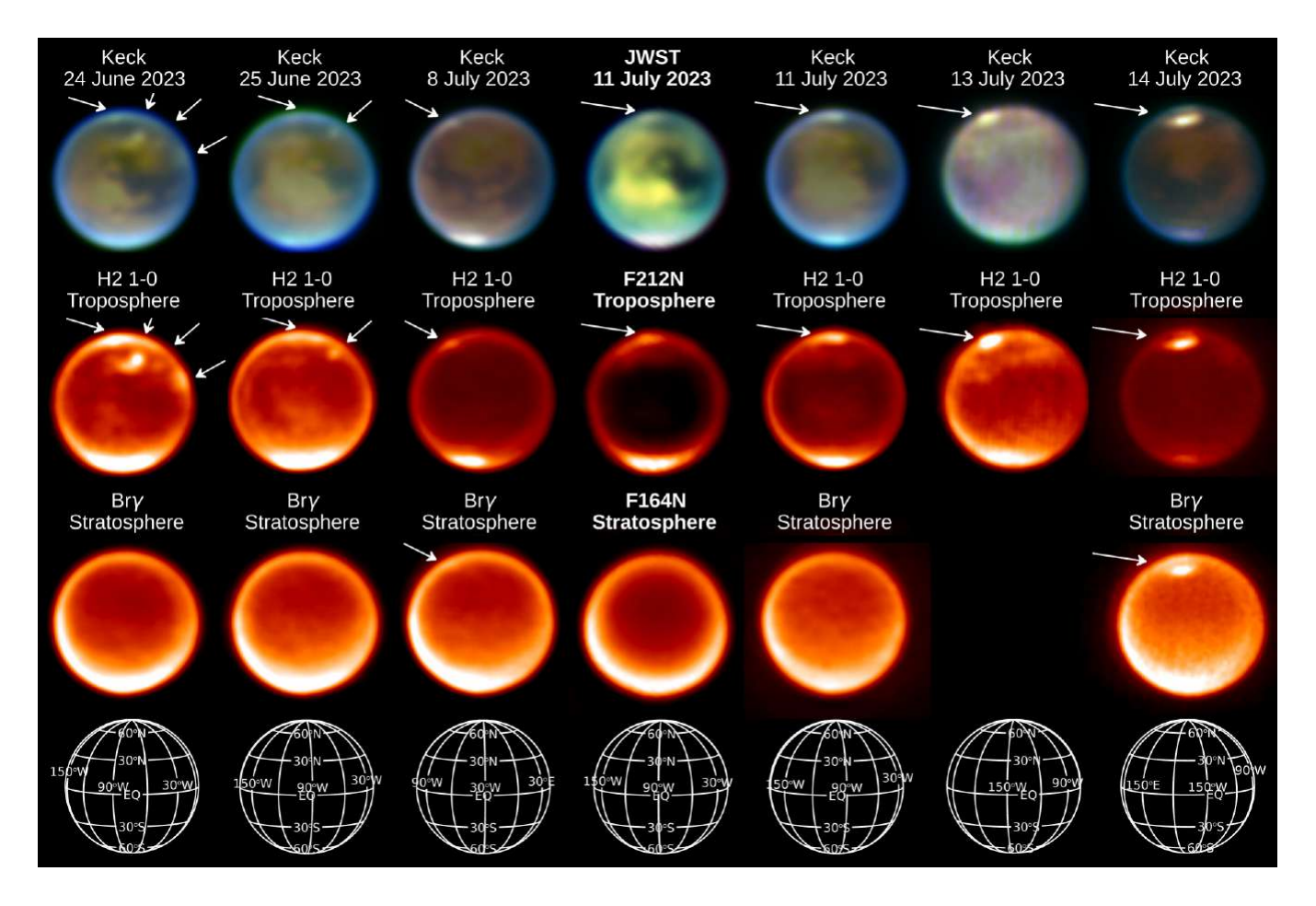}
\vspace{5mm}
\caption{{\bf Time series of JWST NIRCam and Keck II NIRC2 Titan cloud observations for June and July 2023.} The content of each row and image processing follow that of Figure \ref{fig:cloud_images1}a. Note that the NIRC2 Br-$\gamma$ image was not available for the 13th of July, 2023 observation. See \blue{Fig.~\ref{fig:clouds_alts}} for the altitudes probed in the different filters.
}
\label{fig:cloud_images2}
\end{centering}
\end{figure}

\begin{figure}[ht]
\includegraphics[width=1.0\linewidth]{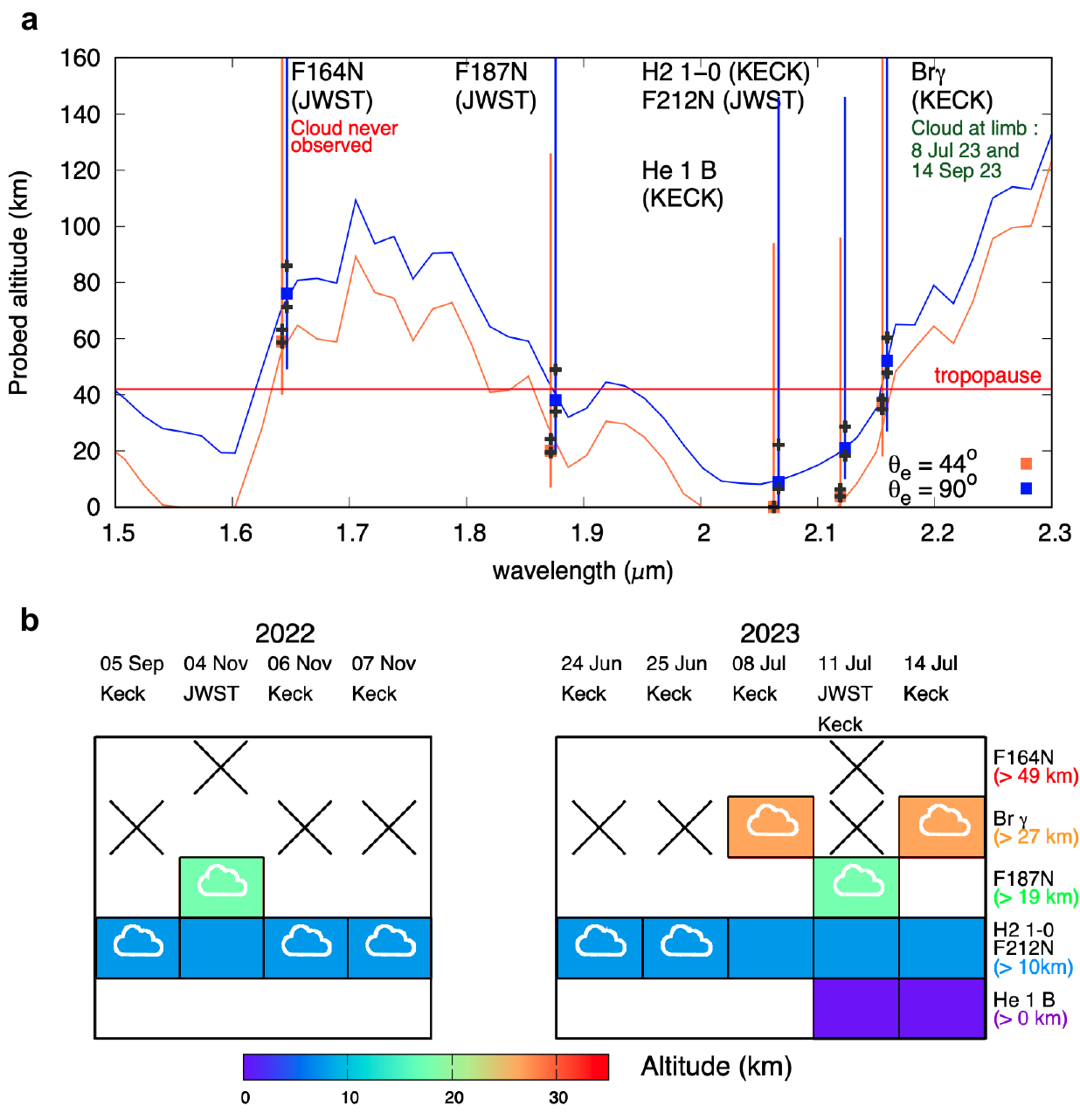}
\caption{
{\bf Altitudes of Titan clouds seen in JWST and Keck imaging.}
{\bf a,} Altitude sensitivity ranges for the NIRCam and NIRC2 filters between 1.5 and 2.3 $\mu$m. We show sensitivity at two emission angles: 
medium emission angle ($\theta_e$ = $44^{\circ}$: orange curves and squares) and limb
($\theta_e$ $\approx 90^{\circ}$: blue curves and squares). Error bars on the altitudes correspond to the region where the effective transmission spans between 10\% and 90\%, thereby encompassing 80\% of the total emerging atmospheric flux (see also \blue{Extended Data Table~1} and \blue{Extended Data Fig.~6}).
The black crosses indicate the sensitivity of the altitude probed for an aerosol opacity increased or decreased by a factor $1.5$. 
{\bf b,} Summary of clouds detected in our observations as a function of time and altitude. Colored boxes indicate clouds detected in a given filter, and boxes with a cloud symbol indicate the highest altitude range at which a cloud was confirmed (`cloud tops'). Black crosses indicate when no cloud is detected in observations, empty (white) squares indicate that no observation was available for the date-filter combination. For each filter, the given altitude is the lowest altitude probed (transmission 10\%) near the limb, for an emission angle of $\simeq 90^{\circ}$. We omit cases with $\theta_e$ = $44^{\circ}$, because clouds are not observed every time and we give no evaluation for the $13^{th}$ July because the single observation in the NIRC2 H2 1-0 filter is not constraining enough (see \blue{Fig.~\ref{fig:cloud_images2}}). 
}
\label{fig:clouds_alts}
\end{figure}

We imaged Titan with JWST's Near Infrared Camera (NIRCam) and the Near Infrared Camera 2 (NIRC2) on the Keck II telescope in late 2022 and mid-2023 using a variety of filters (see \blue{Fig.~\ref{fig:cloud_images1} and \ref{fig:cloud_images2}} and \blue{Supplements A.3} and \blue{A.4}; \blue{Supplements B.3} and \blue{B.4} for reduction). 
Different filters probe to different depths in Titan's atmosphere (see \blue{Methods \ref{sec:methods-nircam-nirc}}, \blue{Extended Data Figures 5 and 6} and \blue{Fig.~\ref{fig:clouds_alts}}) due to varying near-IR opacity primarily due to methane absorption bands, enabling some discrimination of the altitude of observed features.

The first observations of Titan with JWST NIRCam took place on November 4th 2022 (UT) (\blue{Fig.~\ref{fig:cloud_images1}a}). Bright methane clouds were visible near the disk edge in the F212N filter (dawn at left and dusk at right) centered at 60{\dg}N and 50{\dg}N latitude, respectively. We were able to acquire follow-on imaging with the Keck II NIRC2 camera (including adaptive optics) on the nights of November 6th and 7th, which showed clouds at similar latitudes and local times, but with evolving morphologies (\blue{Fig.~\ref{fig:cloud_images1}a}). NIRSpec observations are also available on November 4th 2022 (\blue{Fig.~\ref{fig:cloud_images1}b}), providing a more complete view of the contributions of tropospheric clouds to Titan's near-infrared spectrum. We extracted two spectra from the NIRSpec observations, one on the brightest cloud at the dusk side of the disk and one outside the cloud, at the other side. To make the visual comparison between the two spectra relevant, we choose the off-cloud spectrum to be as close as possible in latitudes and observing geometries to the cloud spectrum, so that the haze contribution and light path length are equivalent for both spectra. This is confirmed in the methane bands, where the two spectra are almost indiscernible. The only few remaining differences can be found in the atmospheric windows (in grey in \blue{Fig.~\ref{fig:cloud_images1}b}), due to the cloud contribution and possible surface albedo differences, and within window wings, particularly brighter for the 2 and 2.7-2.8 windows, due to the cloud contribution only.


Eight months later, on July 11th 2023, we again imaged Titan with JWST NIRCam and with Keck II NIRC2 (\blue{Fig.~\ref{fig:cloud_images2}}); with Keck II we also acquired images on preceding and subsequent nights (see Supplement A.4). 
Clouds were visible at $\sim$70{\dg}N, and appeared to grow in spatial extent and to rapidly evolve in morphology over the week-long period from July 8th to 14th (\blue{Fig.~\ref{fig:cloud_images2}}).


For these two observation sequences, we always detected clouds above 10 km (in the NIRCam F212N or  NIRC2 H2 1-0 filters), corresponding to the mid-troposphere; we detected clouds twice above 19 km (in the NIRCam F187N filter, November 4th 2022 and July 11th 2023) and twice above 27 km (in the NIRC2 Br$\gamma$ filter, July 8th 2023 and July 14th 2023) (\blue{Fig.~\ref{fig:clouds_alts}}; see also \blue{Extended Data Table 1}). 
We did not see clouds in the NIRCam F164N filter (November 4th 2022 and July 11th 2023), which principally probes the stratosphere. Clouds observed above 27 km by Keck were located in the upper troposphere and possibly in the lower stratosphere. Unfortunately, no JWST observations were made on July 8th or 14th in 2023 with filters such as F164N that could confirm with greater certainty that clouds seen by Keck had indeed reached the stratosphere. 

Nevertheless, the overall altitude ranges of the clouds detected with JWST and Keck II, along with their vertical variations on short timescales, suggest tropospheric methane cloud fields developing by convection and rising potentially to the tropopause \cite{griffith2008, barth2010, rafkin2022}. If moist air rises sufficiently and cools, methane readily condenses, leading to cumulus cloud fields that, over the course of a Titan day, can humidify the mid-troposphere and enable rarer deep moist convection extending toward the tropopause \citep{rafkin2022}. Clouds exhibiting such convective behavior were observed in 2004 during late southern summer by Cassini's Imaging Science Subsystem (ISS) over Titan's south pole \citep{Porco2005}, and rising clouds in the southern hemisphere were documented by the spacecraft's Visual and Infrared Mapping Spectrometer (VIMS) \citep{Griffith2005}. Our new observations, demonstrating cloud fields with evolving cloud-top altitudes, for the first time indicate moist convection extending through much of the troposphere in the northern hemisphere. The timing of our observations, corresponding to northern summer, provides new constraints for our understanding of Titan's seasonal weather.  

A schematic representation of the atmospheric circulation in late northern summer is shown in \blue{Extended Data Fig.~7(b)}
During this time, insolation is greatest in the northern hemisphere, and a thermally direct cross-equatorial Hadley circulation sees air rising at northern mid-latitudes and descending in the winter hemisphere \citep{mitchell2006,mitchell2016}. Evaporation from the moist polar surface \citep{faulk2020} leads to moistening of the high-latitude atmosphere, enabling deep moist convection \citep{mitchell2016,battalio2021}; the atmospheric circulation thence transports moisture to lower latitudes \citep{mitchell2016,lora2017,lora2024}.



\clearpage 
\section{Discussion}\label{sect:disc}

With JWST MIRI we have seen the full mid-infrared spectrum of Titan for the first time since the demise of Cassini in 2017, confirming the stratospheric gas emissions seen previously by other mid-IR instruments (Voyager IRIS, ISO, Cassini CIRS, Spitzer IRS). However, the unprecedented sensitivity of JWST has allowed us to detect a very weak emission band of CH$_3$ - the first detection of methyl in the neutral atmosphere. Using a predicted photochemical model profile \citep{19vuietal}, we find the observed emission can be well fitted with either LTE or non-LTE source functions.
However, our calculations show non-LTE limits emission from the thermosphere, so emission originates from the upper stratosphere and mesosphere ($\sim$200--500~km). 

The NIRSpec data from JWST is significantly higher in spectral resolution (by $\sim \times$10) than previously achieved across the full NIR range (1--5 \micron ) by Cassini VIMS, allowing for the first detection of emission from the 4.3-\micron\ band of \coo\ (being in non-LTE in the dayside) and the most detailed measurements yet achieved of the CO non-LTE emission bands near 4.7\,\micron, including the fundamental, first and second hot
bands of $^{12}$CO, and two isotopic bands ($^{13}$CO and C$^{18}$O). 
 Most existing measurements of the CO abundance profile are limited to the lower stratosphere, while VIMS data have recently allowed us to retrieve CO up to about 500\,km \citep[see][and references therein]{fabiano2017}. 
 Our non-LTE forward model radiance calculations, including five CO bands, suggest that CO is well mixed from about 50 km up to about 700 km with a constant value of 55$\pm$5 ppmv, which agrees with photochemical model predictions \citep{19vuietal}.
 
The fact that the \dmethane\ $\nu_2$ emission near 4.54\,\micron\ at disk center is correctly reproduced with a thermal emission model (\blue{Fig.~\ref{fig:co_co2}}) suggests that it is likely in LTE.  This agrees with VIMS measurements that showed that, at a limb tangent height of 350\,km ($\sim$0.01\,mbar), the observed radiance was very similar at daytime and nighttime \citep{fabiano2017}.
This is a sign of negligible solar excitation and henceforth suggests that the observed emission results from LTE, at least up to this altitude. 

The high quality of the JWST spectra offers an unprecedented possibility to address many other interesting topics. These include the
emission of \coo\ at 4.3\,\micron\ (see 
\blue{Fig.~\ref{fig:co_co2}}),
raising the question of why it is in non-LTE while the CH$_3$D $\nu_2$ emission, at a very close wavelength (4.54\,\micron) seems 
to be in LTE. The fact that the \coo\ emission originates at higher altitudes than that of CH$_3$D (see Sec.\,\ref{sec:methods-nirspec}), where we expect non-LTE effects to be more important, might explain the observed features.




Our new observations of methane clouds in Titan's troposphere during late northern summer on Titan add to a catalog of previous detections recorded by ground- and space-based observations \citep{brown2002,roe2002,bouchez2005,schaller2006,schaller2009,Rodriguez2009titan,rodriguez2011titan,turtle2018titan,lemmon2019} that trace out the seasonal variation of Titan's weather over nearly a full year (see \blue{Extended Data Fig.~7}(a)).
In addition to many clouds detected from Cassini, large cloud outbursts detectable from ground-based facilities were common in the southern hemisphere during southern summer into fall, approximately from 2002 to 2011. 
Roughly a year before the northern summer solstice (2017), clouds were increasingly detected in the summer (northern) hemisphere, but few suggested deep moist convection. Our observations indicate a continuation of cloud activity into late northern summer---roughly in agreement with the behavior at high southern latitudes during southern summer---and further suggest the occurrence of deep moist convection extending to the tropopause over the region of Titan where the majority of surface liquids exist. 

The evolving latitudinal pattern of cloud activity with the changing seasons agrees with expectations from recent general circulation model (GCM) simulations \citep{lora2015,mitchell2016,newman2016,faulk2020,lora2022}, as seen in multi-year averages of simulations
(see \blue{Extended Data Fig.~7}(a)).
Simulated precipitation is seasonal and peaks on average over the summer pole; in late northern summer, precipitation is concentrated in northern mid-latitudes. The simulations indicate increased cloud activity in the north at around the equivalent season when large events were observed in the south \citep{schaller2009}, consistent with our observations. Based on the favorable comparison to simulations, we expect continued sporadic, and perhaps decreasing, cloudiness in the northern mid-latitudes, followed by a relatively quick transition of clouds to the southern hemisphere in the late 2020's.



Finally, our findings also demonstrate that cloud evolution in altitude, a qualitative indicator of moist convection and lower tropospheric methane humidity \citep{barth2010}, can be traced using astronomical techniques thanks to an efficient collaboration between observatories (here JWST and Keck). 
Observations of Titan with JWST (and Keck) in Cycles 2 \& 3 (2024 and 2025) will enable further insights into the evolution of clouds in the approach to northern fall equinox (April 2025), a time of active meteorology at low latitudes (large methane and dust storm outbursts as happened shortly after the last equinox in 2010 \cite{turtle2011seasonal, turtle11rapid, rodriguez2011titan, rodriguez2018}). 




\clearpage
\section{Methods}\label{sec:methods}

\subsection{MIRI Modeling} 
\label{sec:mirimodel}


To model the MIRI spectrum in the region of the 16.5~$\mu$m CH$_3$ feature we used the NEMESIS radiative transfer code \cite{irwin2008nemesis}.
NEMESIS has been used extensively to model Titan's spectra in the mid-IR and fitting methods are well tested \cite[][and references therein]{19teaetal}.
Generation of synthetic spectra require the temperature and composition of Titan's atmosphere to be defined.


The temperature profile used to generate the synthetic MIRI spectrum is based on Cassini measurements from half a Titan year ago (2008.8) to capture Titan in a similar seasonal state. The MIRI spectra do not spatially resolve Titan and the resulting spectrum is a disk-average. To account for this we define a nominal temperature profile along with hot and cold end-member profiles 
(see \blue{Extended Data Fig.~2}(a)).
In the troposphere and lower stratosphere ($p>$40~mbar) we use the Huygens descent profile \cite{fulchignoni2005situ}. In the stratosphere and mesosphere (40--0.002~mbar) we use the mean, minimum, and maximum temperature at each pressure level spanning latitudes $\pm$45$^{\circ}$N from Cassini/CIRS in 2008.8, which comprises a compilation of nadir and limb analysis \cite{19teaetal}. In the thermosphere, temperatures are highly variable. For the nominal thermosphere we use a simple isothermal extrapolation of the Cassini/CIRS results for the nominal case, which is consistent with the Huygens entry profile \cite{fulchignoni2005situ} and ground-based results from ALMA \cite{22theetal}. For the cold end-member thermosphere we use a linear extrapolation (in log pressure) from the CIRS cold mesopause value to the coldest measurement of 110~K at $2{\times}10^{-11}$~mbar from Cassini/INMS \cite{13snoetal}. For the hot end-member thermosphere we use a linear extrapolation (in log pressure) from the CIRS hot mesopause value to 200~K at $10^{-6}$~mbar based on Cassini/UVIS stellar occultations \cite{11kosetal} and then 175~K at $2{\times}10^{-11}$~mbar from Cassini/INMS \cite{13snoetal}. 
Note the hot/cold disk-average temperature profile end members used for the MIRI analysis encompass the temperature profile used for the NIRSpec analysis and are consistent with our temperature inversions from JWST observations of 7.7~$\mu$m CH$_4$ emission.


Nominal atmospheric composition and photochemical aerosol profiles are based on equatorial abundances from Cassini/Huygens measurements \cite[full details in][supp. info.]{19teaetal} and aerosol spectral properties are based on Cassini/CIRS \cite{12vinetal}.

Line data are based on GEISA \citep{21deletal} and HITRAN \citep{22goretal} for most atmospheric gases. However, CH$_3$ is not included in these databases, so we use the line list recently applied to Jupiter \cite{20sinetal}, which is based on \cite{98bezetal,99bezetal} with updates to the line intensities from \cite{05staetal}.
We follow \cite{99bezetal} and \cite{96robetal} and assume a Lorentz broadening width of 0.035~cm$^{-1}$atm$^{-1}$ and a temperature dependence coefficient of 0.75.
Collision induced absorption from pairs of N$_2$, CH$_4$, and H$_2$ were including based on the tabulations in \cite{86borfroa,86borfrob,86borfroc,87borfro,91bor,93bortan}.


For efficiency of calculation we used the correlated-$k$ approximation \cite{89gooyun,91lacoin}, which has been extensively used previously for Titan atmospheric modelling \cite[e.g][]{12teaetal,19teaetal,Coy2023}.
Instrumental spectral resolution is incorporated directly into the $k$-tables using the method of \cite{irwin2008nemesis}. Resolution is channel and wavelength-dependent, so a separate instrument function was used for each wavelength \cite{21labetal}.

The $k$-tables were generated on a wavelength grid corresponding to those of the MIRI Titan observations; giving a spacing of 0.0025~$\mu$m in Channel 3 (Long) used for the CH$_3$ analysis. In generating the $k$-tables, absorption coefficients were pre-tabulated on a pressure grid covering 2$\times$10$^{3}$--2$\times$10$^{-9}$~mbar at 30 pressure levels evenly spaced in $\log(p)$ and 15 temperatures evenly spaced from 70--300~K.
Each wavelength, pressure, and temperature point in the $k$-table had absorption coefficients parameterised using 50 $g$-ordinates using a Gauss-Lobatto quadrature scheme \cite{irwin2008nemesis}.

Synthetic MIRI spectra were generated assuming an emission angle of 45$^{\circ}$, which provides a simple approximation to a disk-averaged spectrum.
A second order polynomial fit to the continuum was used to convert the modelled radiances into a line-to-continuum ratio.

The very high abundance of CH$_3$ at low pressure/high altitude means that under an assumption of local thermodynamic equilibrium (LTE) the majority of CH$_3$ emission would originate in the thermosphere, with a contribution function peaking around $10^{-6}$~mbar (800~km)
(see \blue{Extended Data Fig.~2}(c)).
However, at such low pressures LTE is unlikely \cite{02loptay}.

We incorporated non-LTE effects approximately by scaling the usual LTE Planck emission by the function \cite[see, Eq. 5.2, p. 128, in][]{02loptay}, 
\begin{equation}
f(p,T) = \frac{k_1\, {\rm [M]}} { A + k_1\, {\rm [M]}, \label{eq:nlte}}
\end{equation}
where [M] is the total atmospheric number density (molecules\,cm$^{-3}$) at pressure $p$ and temperature $T$ given by [M]=$10^{-6}p/k_{\rm B}T$ where $k_{\rm B}$ is the Boltzmann constant, 
$A$\,=\,3.23~s$^{-1}$ is the Einstein spontaneous emission coefficient of the \ch3(\ni2) band from \cite{05staetal}, $k_1$\,(\cms) is the  rate coefficient of the collisional process $k_1$:\,\ch3(\v2=1)+M\,$\leftrightharpoons$\,\ch3(\v2=0)+M, where M is the colliding molecule, which is most likely to be \n2 or \ch4. 
Note that in the approximation above (Eq.\,\ref{eq:nlte}), the excitation by absorption of photons in the \ni2 band has been neglected. As this band is optically thin, most of the absorbed photons will come from near the tropopause through CIA emission, which occurs at a very low temperature and is thus expected to be negligible. 
Direct excitation by solar flux at 16~$\mu$m is negligible, so is not included.


There are no available measurements for the collisional rates of \ch3(\v2) with \n2 and \ch4. As the \ch4 concentration is about a factor of 50 smaller than that of \n2, and the collisional rate with \n2 is very uncertain (see below), we consider only the collisional relaxation of \ch3(\v2) with \n2.
The rate of this process in collisions with M=He, \k1(He), has been calculated theoretically by \cite{Ma2013} with values of 1.53\e{-12}, 2.06\e{-12}, 2.65\e{-12} and 5.54\e{-12}\,\cms\ at temperatures of 150\,K, 175\,K, 200\,K and 300\,K, respectively. Furthermore, \k1(He) has been measured by \cite{Callear1970} at room temperature obtaining a value of  (8$\pm$2)\e{-13}\,\cms\ that, by applying the temperature dependency of \cite{Ma2013}, is translated in a range of (2.2--3.7)\e{-13}\cms\ at 175\,\,K (a temperature close to that of Titan's thermosphere).
Taking into account both studies, we have for 
\k1(He) a range of (2.2--21)\e{-13}\,\cms\ at 175\,K.

Hovis and Moore \citep{Hovis1980} measured the collisional deactivation of a similar molecule and energy level, \nh3(\v2), \k2, in collisions with He and \n2 at a temperature of 198\,\,K obtaining values of (1.1$\pm$0.2)\e{13}\,\cms\ and (1.9$\pm$0.3)\e{13}\,\cms, respectively.  Later, \cite{Danagher1987} measured the collisional rate of \nh3(\v2) with \n2 at 200\,K obtaining a rate of (1.54$\pm$0.18)\e{-13}\,\cms, slightly lower than that of \cite{Hovis1980}. 
Thus, from these measurements we obtain for \k2(\n2) a range of (1.4--2.2)\e{-13}\,\cms\ at 200\,K.

By taking into account that the deactivation of \nh3(\v2)
with \n2 is $\sim$1.72 larger than for collisions with He \cite{Hovis1980}, and assuming the same dependency for \ch3(\v2), the range of (2.2--21)\e{-13}\,\cms\ obtained above for \k1(He) from \cite{Ma2013} and \cite{Callear1970} is translated into a range of (3.8--34)\e{-13}\,\cms\ for \k1(\n2). Taking into account both the results for the collisions of \ch3(\v2) by He and for \nh3(\v2) by He and \n2, we end up with the largest estimated range for \k1(\n2) of (1.4--34)\e{-13}\,\cms. 
In our study, we have adopted a value of 7\e{-13}\,\cms, which is the geometric mean of the uncertainty range, and assumed a total uncertainty of a factor of 10, i.e., a range of (2.2--22)\e{-13}\,\cms, very similar to that obtained above. 

The non-LTE emission ratio $f(p,T)$ is shown in \blue{Extended Data Fig.~2}(b)
along with the resulting contribution functions spanning our adopted \k1 range 
(\blue{Extended Data Fig.~2}(d-f)).
Non-LTE effects reduce thermospheric emission and contribution functions are predicted to peak around 200--500~km in the stratosphere and mesosphere.
Note, if the actual \k1 rate is smaller than our lower limit, non-LTE effects in the thermosphere would be larger, leading to a smaller radiance and requiring a larger \ch3 concentration in the thermosphere to match the measured radiance.

\subsection{NIRSpec Modeling} \label{sec:methods-nirspec}

We modeled solar reflection and thermal emission through a radiative transfer code based on the radiative solver DISORT (version 2.0) \cite{Stamnes1988}. We included gas opacity from methane ($^{12}$CH$_4$, $^{13}$CH$_4$ and \dmethane), CO, \coo, \acet, HCN, \ethane\ and N$_2$-N$_2$ pairs.  Spectroscopic data are based on HITRAN \citep{22goretal} except for \methane\ \citep{Rey2018} and \ethane\ \citep{Hewett2020}. We used the correlated-$k$ approximation in the radiative transfer \citep{89gooyun,91lacoin}. For each NIRSpec wavelength and each species, a set of 8 $k$-coefficients were used: 4 for the interval [0:0.95] of the normalized frequency $g$ and 4 for the interval [0.95:1.0]. These were first calculated from a line-by-line radiative transfer code for a set of pressures and temperatures that encompass Titan's atmospheric conditions up to 600\,km. The wavelength-dependent spectral resolution of NIRSpec was taken into account at this stage. In the radiative transfer, the $k$-coefficients of the different species are combined assuming no correlation between species as described in \cite{91lacoin}. The \methane\ profile used is that derived from Huygens/GCMS measurements \citep{niemann2010composition} while those from the other gases and the D/H ratio are based on Cassini/CIRS measurements at low latitudes \citep{Mathe2020, Bezard2007}. The CO mole fraction is held constant at 55\,ppmv, a value that best reproduces the NIRSpec data beyond 4.5\,\um. We used the haze vertical opacity profile and phase functions modeled from Huygens/DISR in situ measurements \citep{Doose2016}. The aerosol single scattering albedo and surface (Lambertian) albedo were adjusted at a few reference wavelengths: in the core of the \methane\ bands and in the transparency windows, following \citep{Hirtzig2013}. 

Thermal emission cannot be neglected compared with solar reflection beyond $\sim$3.7\,\um. The temperature profile used in the radiative transfer model was based on Cassini/CIRS measurements near 11\dg S from half a Titan year ago (November 2008) \citep{Mathe2020} to represent approximately the seasonal state observed by JWST at disk center (14\dg N). However, calculations incorporating this temperature profile overestimate the \dmethane\ emission in the range 4.3--4.6\,\um\ and produce excess haze emission in the troughs around 4.7\,\um. This discrepancy likely results from the asymmetry of Titan's seasons as a result of Saturn's orbital eccentricity. We then constructed a cooler temperature profile following the predictions of a simple seasonal radiative model \citep{Bezard2018}. 
The resulting temperatures are cooler above the 15-mbar level by 3.3\,K at 4 mbar (130 km) and 4.5\,K at 0.2\,mbar (260 km), varying linearly with log-pressure.
With this ``nominal'' temperature profile, thermal emission from \dmethane\ 
mostly originates from 170--270 km and satisfactorily reproduces the observed emission line intensities to within 15\%. \dmethane\ emission also depends on the methane mole fraction in the stratosphere. Using a lower value of 1\%, as derived at some latitudes from Cassini/CIRS measurements \citep{lellouch2014} and from Huygens/DISR in situ measurements \citep{Rey2018}, the derived temperature profile is warmer than the nominal one by 2.2 K at 4 mbar and 3 K at 0.2 mbar. However, this ``warm'' temperature profile 
results in a mismatch in the relative line intensity between the fundamental and hot bands of CO (see below).

The calculation of the non-LTE radiance requires first the computation of the non-LTE population of the emitting levels. 
The \coo\ non-LTE model was adapted from the GRANADA model for the Earth \cite{Funke2012}. 
As GRANADA is a generic model, this implementation consists mainly in adapting the collisional processes and optimizing the solution of the coupled system of the statistical equilibrium equations for the considered \coo\ vibrational levels and of the radiative transfer equations for the transitions between those levels. The more important changes are summarized below. As for Earth, the \coo\ energy levels where coupled with \n2(1), which, in the case of Titan, is  also coupled with the CO and \ch4 vibrational levels (see Fig.\,4 in \cite{fabiano2017}). The collisional processes that affect \n2(1) and their rate coefficients are described in \cite{fabiano2017} and those included for the vibration levels of \coo\ are listed in \cite{Funke2012}, with the most recent values for the exchange between \coo($\nu_3$) and \n2(1) derived by \cite{Jurado-Navarro2015}. The solution of the system of equations is carried out as described in \cite{Funke2012} but simplified for the transitions arising from the energy levels with $\nu_2>3$.
The calculations were carried out  using the temperature profile discussed above and a constant CO$_2$ vmr of 18\,ppbv. The illumination conditions were those prevailing during the measurements: a Sun-Titan distance of 9.846 AU, a mean solar zenith angle (SZA) of 22\dg, and an emission angle of 20\dg.

The CO non-LTE model previously developed for the analysis of Cassini VIMS data \cite{fabiano2017} was used here, but it was updated in two aspects. First, the collisional exchange between CO($v$=2) and \nitrogen($v$=1) was increased by a factor of two. Otherwise, the NIRSpec emission in the lines of first hot band of CO was overestimated. This only affects the population of CO($v$=2) below about 200\,km and varies in only a very few K (see \blue{Extended Data Fig.~3}.
We should note that VIMS limb measurements were not sensitive to the emission from the CO($v$=2) below about 200~km because of the strong scattering contribution. Hence, this result represents an improvement of the non-LTE model previously developed. 
Secondly, the model has been extended to calculate the population of the CO($v$=3) level in order to compute the emission  of the second hot band CO(3$\rightarrow$2) detected by NIRSpec. The CO($v$=3) has been considered excited by solar absorption in the overtone CO(3$\rightarrow$0) band near 1.55\,\um, and coupled by vibrational-vibrational collisions to the rest of the CO and \nitrogen\ levels analogously as for CO($v$=2). The calculations were carried out using the nominal temperature profile discussed above, a constant CO vmr of 55\,ppmv and the illumination conditions detailed above for CO$_2$. The CO($v$=1) level of the most abundant isotopologue is in non-LTE above around 300\,km while the rest of the levels are in non-LTE in virtually the entire atmosphere. 

The nadir non-LTE radiances of CO and \coo\ 
were carried out using the Karlsruhe Optimized and Precise Radiative Transfer Algorithm \citep[KOPRA;][]{Stiller2002} and including the non-LTE populations described above. 
The spectroscopic data for CO and \coo\ were taken from the HITRAN 2016 and 2012 compilations, respectively \citep{Gordon2017,Rothman2013}. 
The NIRSpec instrument line shape with a resolving power close to 2700 \citep{Jakobsen2022} was included in the calculations. 

To find out which atmospheric regions we are sensing with each CO band, we calculate the contribution functions of the five bands 
(see \blue{Extended Data Fig.~4})
where the two minor isotopic bands have been grouped). The probed regions are the result of the combined effects of the opacity of the ro-vibrational lines and their vibrational temperatures. 
We see in \blue{Extended Data Fig.~4})
that the strongest fundamental band of the main isotopologue contributes from $\sim$50\,km to near $\sim$700\,km, with two peaks near $\sim$200\,km and $\sim$500\,km. The lower peak occurs in the middle stratosphere, where this band is in LTE and the emission takes place in the moderate lines. The upper peak corresponds to a region of non-LTE emission, where the population of CO (1) is much larger than that of LTE (see \blue{Extended Data Fig.~3}),
and the emission originates mainly from the core of the stronger lines of the R and P branches. The first hot band (FH) also shows a double-peak structure but, as this band is weaker, the emission arises from lower altitudes, in non-LTE conditions.
The isotopic bands also show contributions from two major regions, peaking near 100\,km and 275\,km. The weaker second hot band contributes mainly in the lower troposphere and is also principally in non-LTE. 
In general, we see that the JWST measurements probe the CO concentration from the upper troposphere ($\sim$50\,km) all the way through the lower thermosphere ($\sim$700\,km).

We also tested the sensitivity of the CO emission to the kinetic temperature. Using the ``warm'' temperature profile instead of the nominal one increases the line intensities of the fundamental CO(1$\rightarrow$0) band by around 10\,\%, while those of the hot bands remain unaffected. The relative intensities of the different bands are therefore not as well reproduced as with the nominal temperature profile. Consequently, we used mainly the intensities of the hot bands to derive our best-fit CO mole fraction because they are almost insensitive to uncertainties in stratospheric kinetic temperatures.

\subsection{Modeling of the altitudes probed by JWST/NIRCam and Keck/NIRC2 filters} \label{sec:methods-nircam-nirc}

\noindent

Gas and particle absorption and scattering properties are the main 
components that control the way the solar flux penetrates in Titan's atmosphere. 
Intensities observed in a given channel (JWST/NIRSpec) or filter (JWST/NIRCam and Keck/NIRC2) are generally modeled by accounting for the sub-resolution spectral variations of gas properties, either with line-by-line calculations, with a correlated-$k$ 
approach (as we do here), or using an exponential sum fitting method, or other means. In this work, particle and gas properties are calculated following the method of 
\cite{Coutelier2021}.

The penetration depth can be defined as the altitude where the flux transmission reaches a value of $e^{-1}$. Within a channel, the effective flux transmission is given by $T_{\rm eff}=\sum_{i=1}^{N} w_i T^*_i$ where $w_i$ are the weighting coefficients and $T^*_i$ the effective flux transmission for the $i^{th}$ term of the correlated-k method. This $i^{th}$ term of the effective flux transmission $T^*_i$ is given by $e^{-\tau^*_i}$ where $\tau^*_i$ is the integral along the line of sight of the quantity $d\tau^*_i(z) = \left[ (1-\varpi_i(z))(1-\varpi_i(z) g_i(z))\right]^{1/2} \times d\tau_i(z)$ 
(\cite{Pollack1985}) and $\varpi_i(z)$, $g_i(z)$ and $d\tau_i(z)$ are the average single scattering coefficient, the average asymmetry parameter and the optical thickness of the discrete layer at the altitude  $z$ for the correlated term $i$ respectively.

This method is especially suitable to account for the path of the light in scattering media. We then write the average altitude of penetration as $T_{\rm eff}(z^*)=\sum_{i=1}^{N} w_i e^{-\tau^*_i}=e^{-1}$ where $\tau^*_i$ can be easily evaluated from any model matching data. This altitude also corresponds to the level where $\tau_{\rm eff} = - \log(T(z^*)) = 1 $ where $\tau_{\rm eff}$ is the effective opacity through a given NIRSpec channel or a NIRCam or Keck/NIRC2 filter. Using NIRSpec resolution, we obtain the altitude probed $z^*$ shown in \blue{Extended Data Fig.~5}
at three incidence angles, showing that one can probe Titan's atmosphere from the surface in the most transparent windows, to above $200$ km in the most absorbing band. From the transmission curves (\blue{Extended Data Fig.~6})
we also find the centre and width of the probed region by the JWST/NIRCam and Keck/NIRC2 filters by setting the altitudes where $T(z^{*})=e^{-1}$ for the centre, $T(z_{high})=0.9$ for the higher altitude and  $T(z_{low})=0.1$ for the lower altitude. The latter is the threshold altitude for the detection of a cloud in a given filter. This is reported in \blue{Extended Data Fig.~5} and \blue{ Extended Data Table~1}.

The observations of Titan in different filters make it possible to give constraints on the cloud top altitude on different observation dates (see \blue{Fig.~\ref{fig:clouds_alts}} and \blue{Extended Data Table~1}.
For this, we compare the threshold altitudes corresponding to the filter where clouds are observed, either at the limb with emergence angle $\theta_e=90$\dg or at an emergence angle of  $\theta_e=44$\dg, which corresponds to the areas least distant from the center where the clouds are observed (\blue{Fig.~\ref{fig:cloud_images1} and \ref{fig:cloud_images2}}). When a cloud is detected simultaneously in several filters, we conclude that the cloud extends to an altitude above the highest possible threshold level consistent with observations. This gives a constraint on the minimum cloud top altitude. (Our analysis does not allow for characterizing the base altitude of the cloud.) These altitudes are shown with a cloud icon in \blue{Fig.~\ref{fig:clouds_alts}}. 


\clearpage 

\backmatter

\clearpage

\clearpage 
\section*{Declarations}

\bmhead{Data availability}

JWST observational data is accessible from the Barbara A. Mikulski Archive for Space Telescopes (MAST) from {\tt https://mast.stsci.edu}.
All Keck data, including the Twilight Zone data, are public after 18 months, and can be retrieved from the Keck Observatory Archive (KOA): \url{https://koa.ipac.caltech.edu/UserGuide/about.html}. 

\bmhead{Code availability}

The custom pipeline and data processing code \citep{King_2023} used in this study is available from {\tt https://doi.org/10.3847/2515-5172/ad045f}.
NIRC2 distortion files and IDL software to correct the images for distortion can be downloaded from \url{https://www2.keck.hawaii.edu/inst/nirc2/dewarp.html}. General software to reduce Keck NIRC2 (Twilight Zone) data can be obtained from \url{https://nirc2-reduce.readthedocs.io/en/latest/}.
The NEMESIS modelling software used for MIRI analysis is fully described in  \citep{irwin2008nemesis} and available from {\tt https://github.com/nemesiscode/radtrancode}.  The radiative transfer code used to generate the NIRSpec LTE spectra is available from B.\ B\'ezard upon reasonable request.
The KOPRA radiative transfer code and the GRANADA non-LTE code used in the analysis of the CO$_2$ and CO emissions are available from M. López-Puertas upon request. 

\bmhead{Acknowledgments}

CAN was funded for this work by JWST Archive Research Project \#02524. BB, EL, PR, SR and MES acknowledge support from the Programme National de Plan\'etologie (PNP) of CNRS-INSU co-funded by CNES. SR also acknowledges financial support from the CNES and the French National Research Agency (ANR-21-CE49-0020-04/RAD$^3$-NET and ANR-23-CE56-0008/EOLE). M.L.-P. acknowledges financial support from the Agencia Estatal de Investigaci\'on, MCIN/AEI/10.13039/501100011033, through grants PID2022-141216NB-I00 and CEX2021-001131-S. NAT was funded by UK Science and Technology Facilities Council (STFC) grant ST/Y000676/1. NAL and JML were funded by NASA CDAP Grant 80NSSC20K0483.

This work is based [in part] on observations made with the NASA/ESA/CSA James Webb Space Telescope. The data were obtained from the Mikulski Archive for Space Telescopes at the Space Telescope Science Institute, which is operated by the Association of Universities for Research in Astronomy, Inc., under NASA contract NAS 5-03127 for JWST. These observations are associated with program \#1251.
The authors wish to thank the following staff at Space Telescope Institute for support with the execution of the JWST observations: Katherine Murray, Bryan Hilbert, Glenn Wahlgren, and Blair Porterfield.

Some of the data presented in this paper were obtained at the W.M. Keck
Observatory, which is operated as a scientific partnership among the California Institute of Technology, the University
of California, and the National Aeronautics and Space
Administration. Observing time for this project was allocated by all three institutions, in part thanks to the Twilight Zone Program. NASA Keck time is administered by the NASA Exoplanet Science Institute. The Observatory was made possible by the generous financial support of the W. M. Keck Foundation. The authors wish to recognize and acknowledge the very significant cultural role and reverence that the summit of Maunakea has always had within the indigenous Hawaiian community. We are most fortunate to have the opportunity to conduct observations from this mountain.

\bmhead{Author contributions}

CAN was the Principal Investigator of the JWST GTO 1251 observations analyzed in this paper, led the  team to analyze the data and formulate the results and conclusions, and was also the lead author of this article, writing and editing significant portions of the final manuscript. All other authors indicated with $^\dagger$ on the first page contributed significantly to the data analysis and writing of the paper - other contributions are now described. RKA and AGH contributed to the formulation of the observing proposal GTO 1251 and were co-investigators on the proposal, and also contributed to the discussion of results in this paper. HBH was the lead scientist of the solar system GTO program, encompassing GTO 1251 (Titan) and contributed to the observing proposal and interpretation of the results. BB, EL and MLP were members of the GTO proposal team, and also jointly conducted the modeling of the NIRSpec data (\blue{Fig.~\ref{fig:co_co2}}). YNP contributed with the non-LTE modelling of NIRSpec data (\blue{Fig.~\ref{fig:co_co2}}). TC was a member of the GTO 1251 observing team and involved in the calibration, quality control and navigation of the returned data. NAT was a member of the GTO 1251 proposal team and also conducted analysis and modeling of the MIRI MRS data (\blue{Fig.~\ref{fig:miri-figure}}). BPC and MES calibrated and plotted the spectral data from NIRSpec and MIRI (\blue{Fig.~\ref{fig:combined-spectrum}}), and made assessments of noise level. SR, CS, EPT and RAW were members of the GTO 1251 proposal team and involved with the interpretation of the NIRCam data; SR and JS made images of the NIRCam and NIRC2 data (\blue{Figs.~\ref{fig:cloud_images1} and \ref{fig:cloud_images2}}). SNM and JAS provided guidance on the formulation of the observing proposal GTO 1251, including observational parameters. NRG, BJH, ORK, LNF and MTR contributed to the calibration of the spectra, especially non-standard pipeline processing to improve calibration for bright/extended objects. SCRR, NWK, JML and NAL contributed to interpretation of Titan cloud patterns in comparison to models; in addition JML made \blue{Extended Data Fig.~7}.
PR modeled the light penetration depth in different NIRCam and NIRC filters and made \blue{Fig.~\ref{fig:clouds_alts}}. IdP and KdK were the leads for the Keck observations; IdP observed, reduced and calibrated the data with assistance of TM and EM, and contributed description sections to the paper. JOM and CJ assisted with the scheduling of Keck observations;  CA, GD, AR provided observing support. JOD, CS, HM, LM and TSS wrote a competed observing proposal for Keck time that provided observing time for this project. IdP, KdK, and AGD wrote competed proposals to the three institutions for the Twilight Zone Program, which provided some of the Keck data.

\bmhead{Competing interests}

All authors declare no competing interests.

\bmhead{Ethics approval} 

The authors declare that appropriate ethical guidelines for scientific research and integrity were followed.

\clearpage

\section*{Figure Captions}

\clearpage
\noindent
\large 
\begin{center}
{\Large Extended Data Figures and Tables for Nature Astronomy manuscript NATASTRON-23050477A}
\end{center}
\normalsize

\setcounter{figure}{0}
\renewcommand{\figurename}{} 
\renewcommand{\thefigure}{Extended Data Fig.~\arabic{figure}}

\setcounter{table}{0}
\renewcommand{\tablename}{} 
\renewcommand{\thetable}{Extended Data Table~\arabic{table}}

~~~\\


\begin{table}[ht]
    \centering
    \begin{tabular}{llll}
        \hline
        Telescope & Filter & Slant emergence            &   Limb \\
                  &        & ($\theta_e = 44^{\circ})$  &  ($\theta_e = 90^{\circ}$)\\
        \hline
        JWST &F164N         & 59 [40,205] & 76 [49,227]\\
        JWST &F187N         & 20 [7,126]  & 38 [19,163]\\
        JWST &F212N         & 4 [0,96]    & 21 [10,146] \\
        Keck & He 1B        & 0 [0,94]    & 9  [0,146]\\
        Keck & H2 1-0       & 1 [0,94]    & 19 [9,145] \\
        Keck &  Br$\gamma $  & 36 [18,168 ] &52 [27,195 ] \\
         \hline
        & & & \\
    \end{tabular}
    \caption{{\bf Altitudes in Titan's atmosphere probed by  JWST NIRCam and Keck NIRC2.} These were computed 
    from consideration of the effective flux transmission $T_{\rm eff} = e^{-1} $ at two emergent angles $\theta_e$ for several key filters. The altitudes in brackets (km) are those where the effective transmission $T_{\rm eff}=0.1$ and $T_{\rm eff}=0.9$, giving the threshold and the top altitudes ($z_{\rm low}$ and $z_{\rm high}$ respectively) probed by the filter. This range shows the region where about $80\%$ of the light flux is reflected to the observer or absorbed.}
    \label{tab:nircam_prob_alt}
\end{table}


\begin{figure}[ht]\
\begin{centering}
\includegraphics[width=1.0\linewidth]{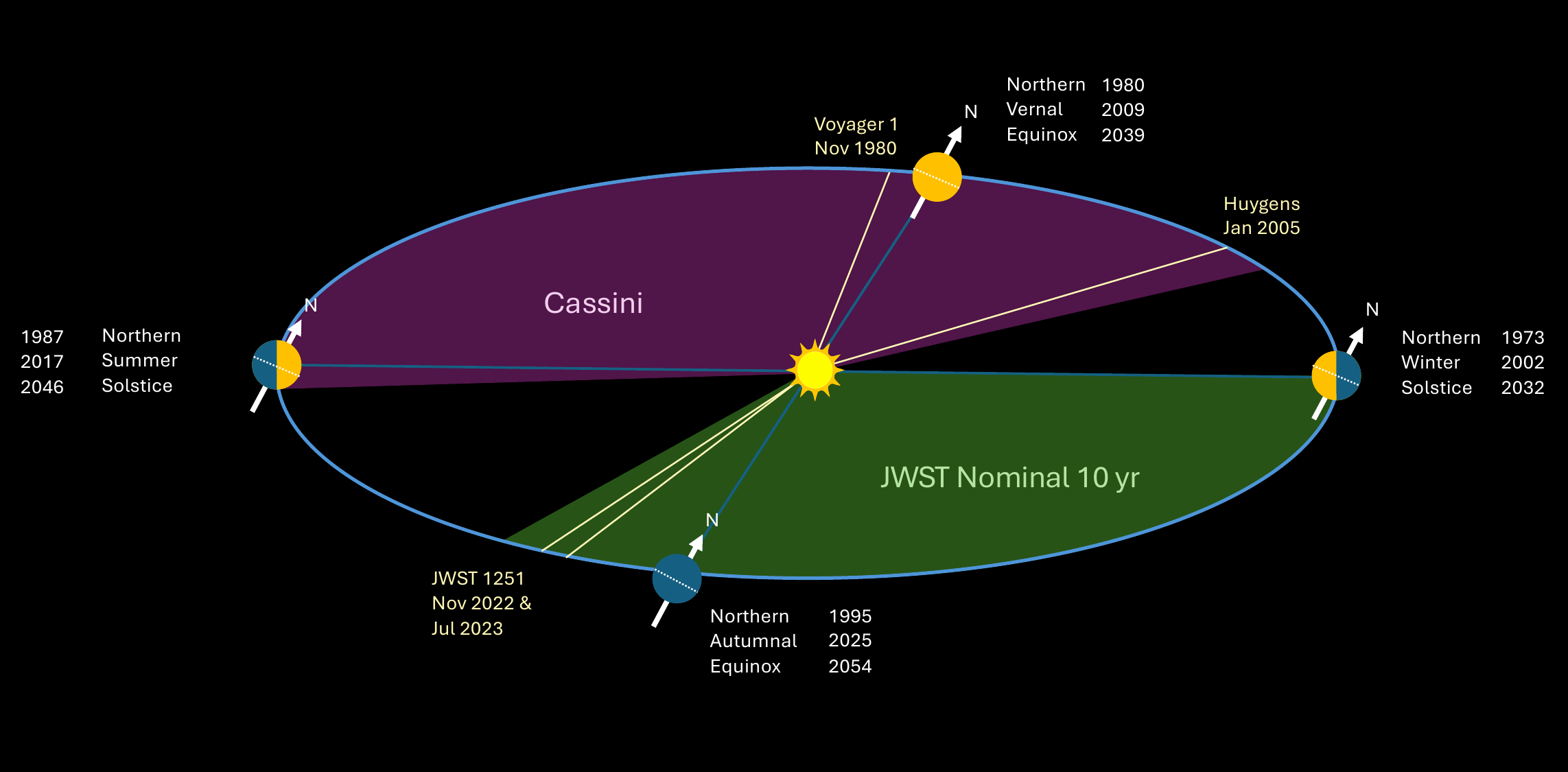}
\caption{{\bf Seasons on Titan seen by various spacecraft missions.} 
Extended observations by Cassini (June 2004 to September 2017) may be compared to JWST (from November 2022). 
The early JWST observations probe a significantly different season to Cassini (late northern summer), last seen in the early 1990s.}
\label{fig:titan_overview}
\end{centering}
\end{figure}


\begin{figure}[!ht]
    \centering
    \includegraphics[width=1.0\linewidth] {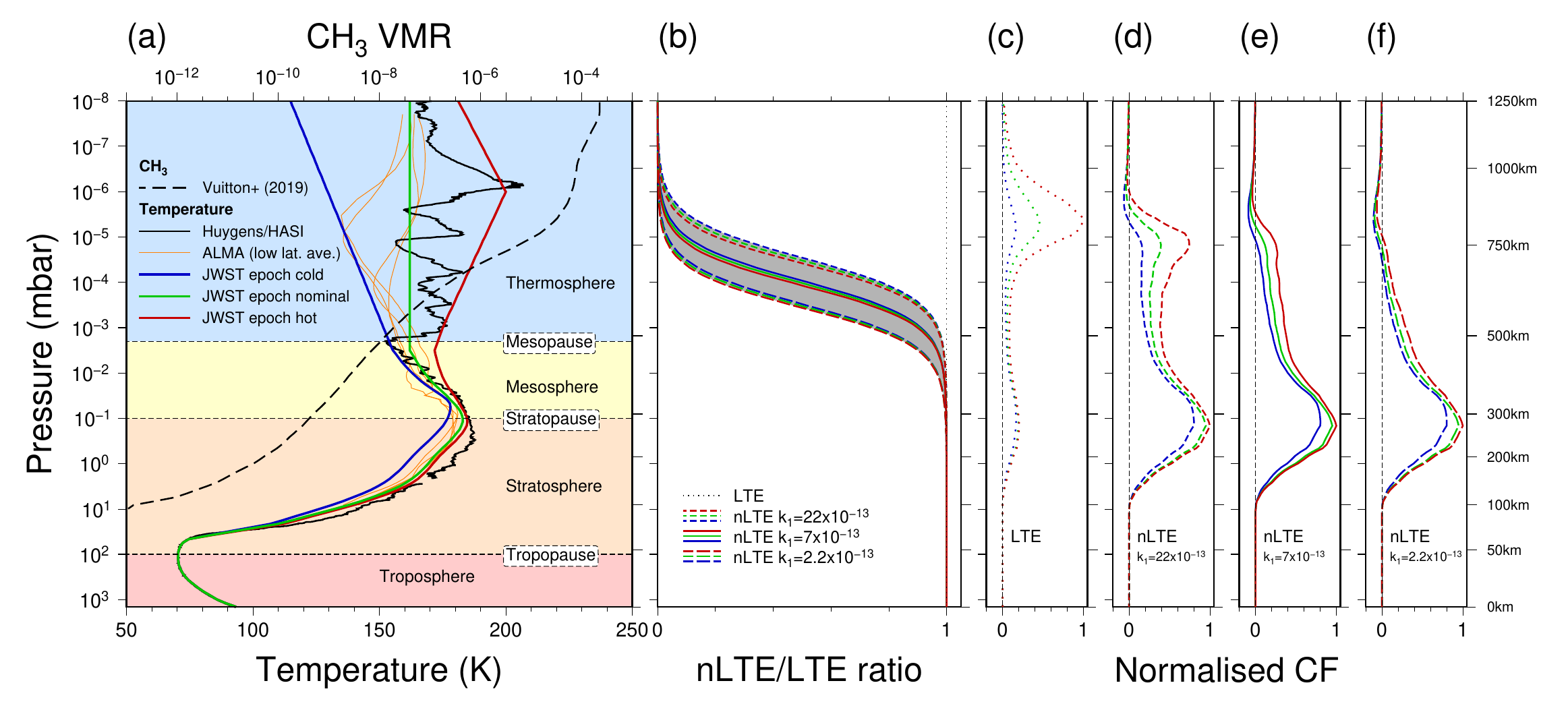}
    \caption{
    {\bf MIRI spectroscopic modeling inputs for the CH$_3$ analysis.} (a) Nominal and end-member temperature profiles and CH$_3$ profile. Stratospheric and mesospheric temperatures are based on Cassini CIRS temperature inversions from 2008.8 (half a Titan year earlier for seasonal correspondence in the latitude range $\pm$45$^{\circ}$N). Thermospheric temperatures are highly variable and end-member profiles are based on low latitude results from Cassini UVIS and INMS, the Huygens probe entry profile at 15$^{\circ}$S, and ALMA equatorial observations. b) non-LTE emission ratio for the range of k$_1$ values considered. The grey region shows the uncertainty in the non-LTE effect and the vertical dotted line shows the LTE case for comparison. (c--f) Normalised contribution functions ($dR/dx$ where $R$ is radiance and $x$ is log(VMR)) for the LTE and non-LTE cases). The non-LTE effect suppresses the thermospheric emission peak so most emission is from the stratopause region in the nominal temperature and k$_1$=7$\times$10$^{-13}$~cm$^3$\:s$^{-1}$ case. Note the slightly negative contribution functions in the thermosphere originate from weak CH$_3$ absorption due to the high abundance, combined with negligible non-LTE emission at those altitudes. 
    }
    \label{fig:miri-tp}
\end{figure}


\begin{figure}[!ht]
\centering
 \rotatebox{90}{\includegraphics[width=0.6\textwidth]{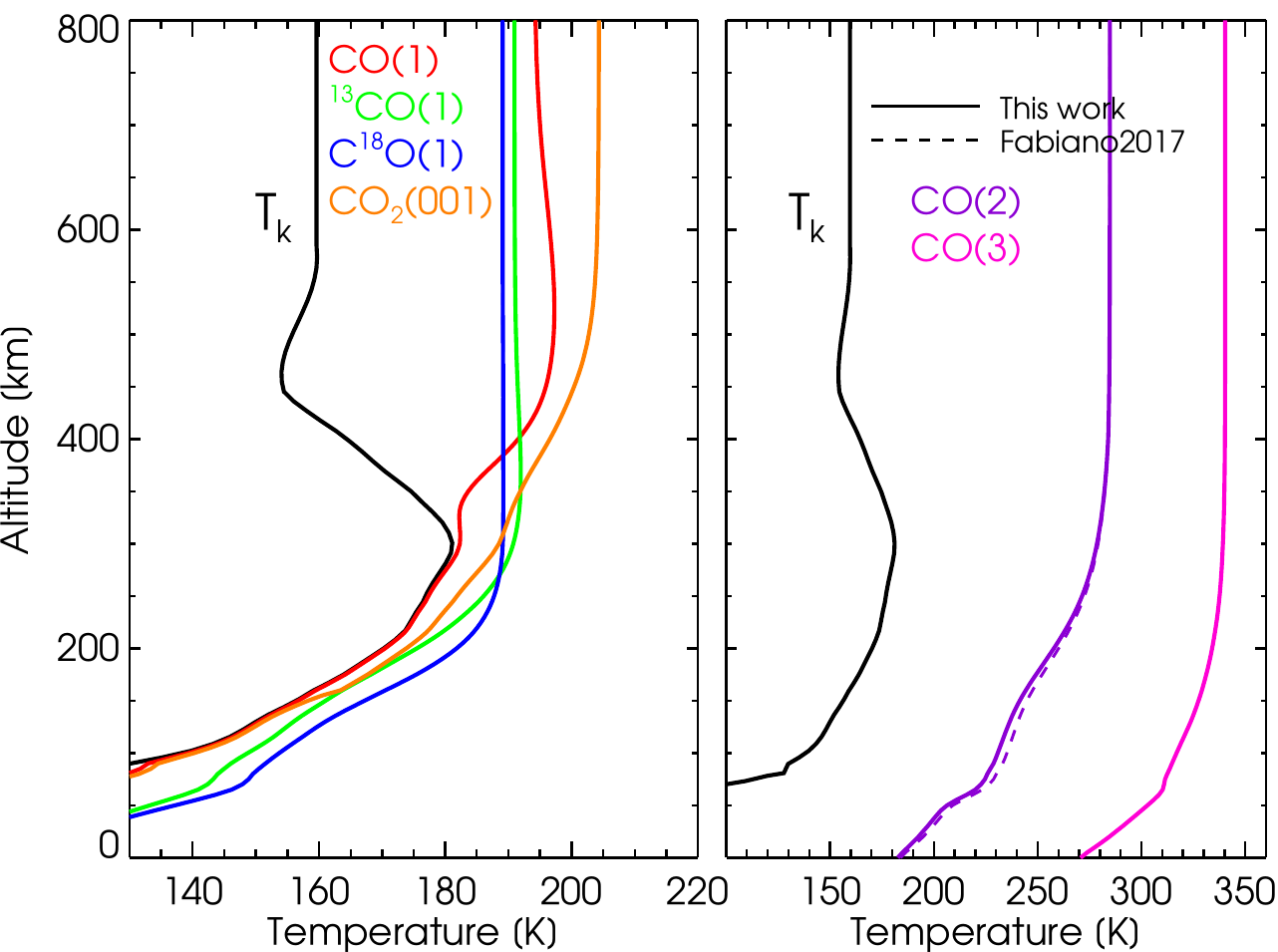}}\\[0.2cm]
 \caption{{\bf Nominal kinetic temperature profile and vibrational temperature profiles for CO and \coo .}
 Left: nominal kinetic temperature ($T_k$, black) profile and  vibrational temperature profiles for the fundamental bands of the CO isotopologues (colors) computed by using the non-LTE model. Right: the vibrational temperatures of the higher energy levels, CO(2) and CO(3), and also that of CO(2) calculated by  \cite{fabiano2017}. 
 }
 \label{fig:co_vts}
 \end{figure}


\begin{figure}[!ht]
\centering
\rotatebox{90}{\includegraphics[width=0.7\textwidth]{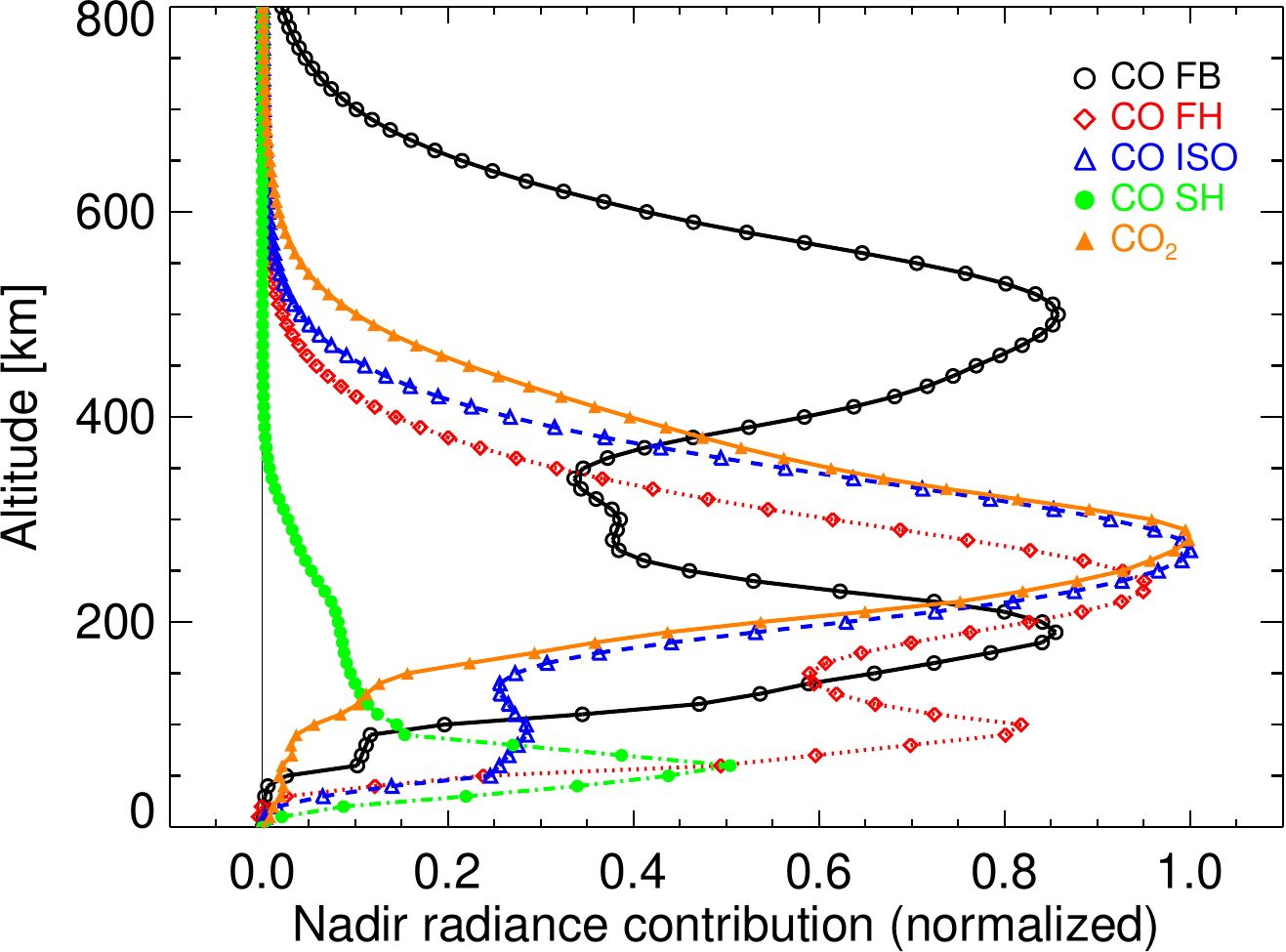}} \\[0.2cm]
\caption{{\bf Contribution to the nadir radiances for the different bands of CO and the $\nu_3$ fundamental band of \coo .} CO bands: FB = Fundamental Band; FH = First Hot Band; ISO = Isotopic  ($^{13}$CO(1$\rightarrow$0) and C$^{18}$O(1$\rightarrow$0)) Bands; SH = Second Hot Band). Contribution function amplitude indicates the atmospheric regions where emission is arising. The radiances have been integrated  in the 4.45--5.00\,\um\ spectral interval for the CO bands and in the 4.20--4.35\,\um\ range for CO$_2$.}
\label{fig:co_weights}
\end{figure}


\begin{figure}[!ht]
\centering
\includegraphics[width=1.0\textwidth]{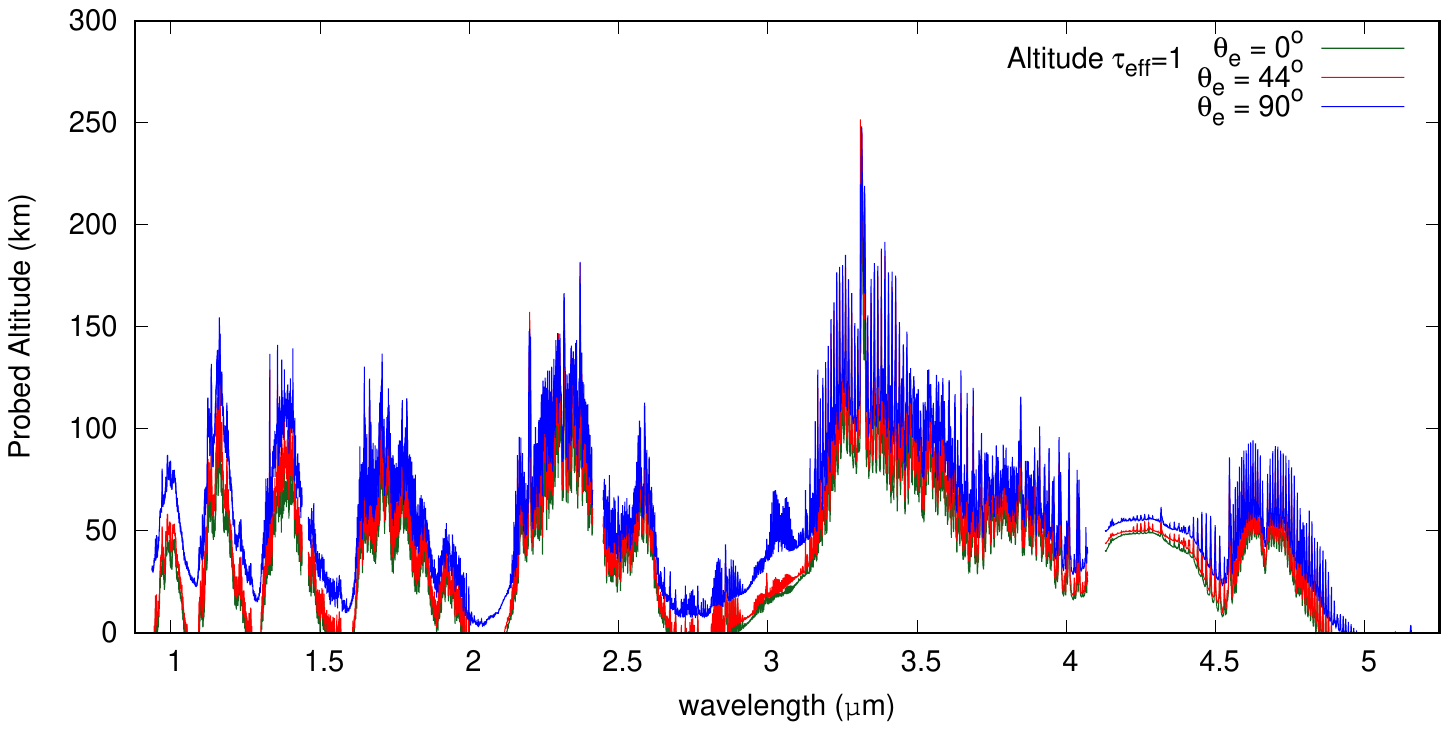}\\[-0.6cm]
\caption{{\bf Penetration depth of the direct and scattered flux for NIRSpec as a function of wavelength.}
Calculations at three emission angles:
$\theta_e  =0^{\circ}$ (nadir viewing, green), $\theta_e = 44^{\circ}$ (red) and $\theta_e = 90^{\circ}$ 
(limb viewing, blue).}
\label{fig:gas-opacities}
\end{figure}


\begin{figure}[!ht]
\centering
\includegraphics[width=1.0\textwidth]{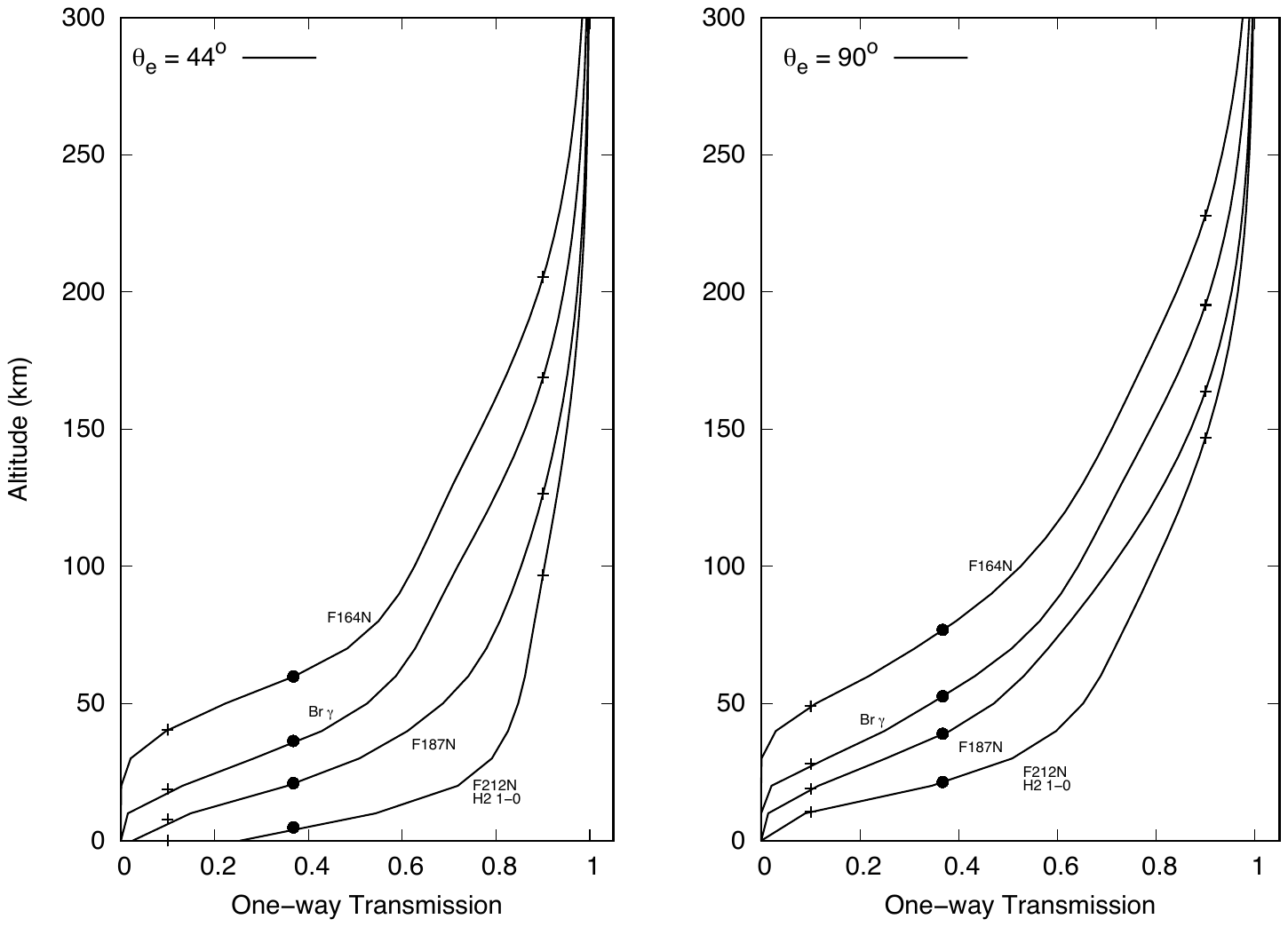}\\[-0.6cm]
\caption{ {\bf One-way transmission through Titan's atmosphere.} Calculations at emission angles $\theta_e=44^{\circ}$ (left) and at the limb $\theta_e=90^{\circ}$ (right) 
for light at different wavelengths probed by JWST NIRCam and Keck NIRC2. JWST filters (wavelengths): 
F165N (1.65 $\mu$m); F187N (1.87 $\mu$m); F212N ($\simeq$ 2.12 $\mu$m). Keck NIRC2 filters (wavelengths):
Br$\gamma$ filter (2.157 $\mu$m); H$_2$ 1--0 ($\simeq$ 2.12 $\mu$m). 
The dots indicate for each filter and viewing orientation the altitude where the effective transmission reaches $e^{-1}$, and on each curve, the crosses indicate the altitude where the one-way effective transmission reaches $90\%$ (upper cross) and $10\%$ (lower cross). 
The latter altitude is also used as the threshold altitude of detection.}
\label{fig:emission-angles}
\end{figure}


\begin{figure}[ht]
\includegraphics[width=1.0\linewidth]{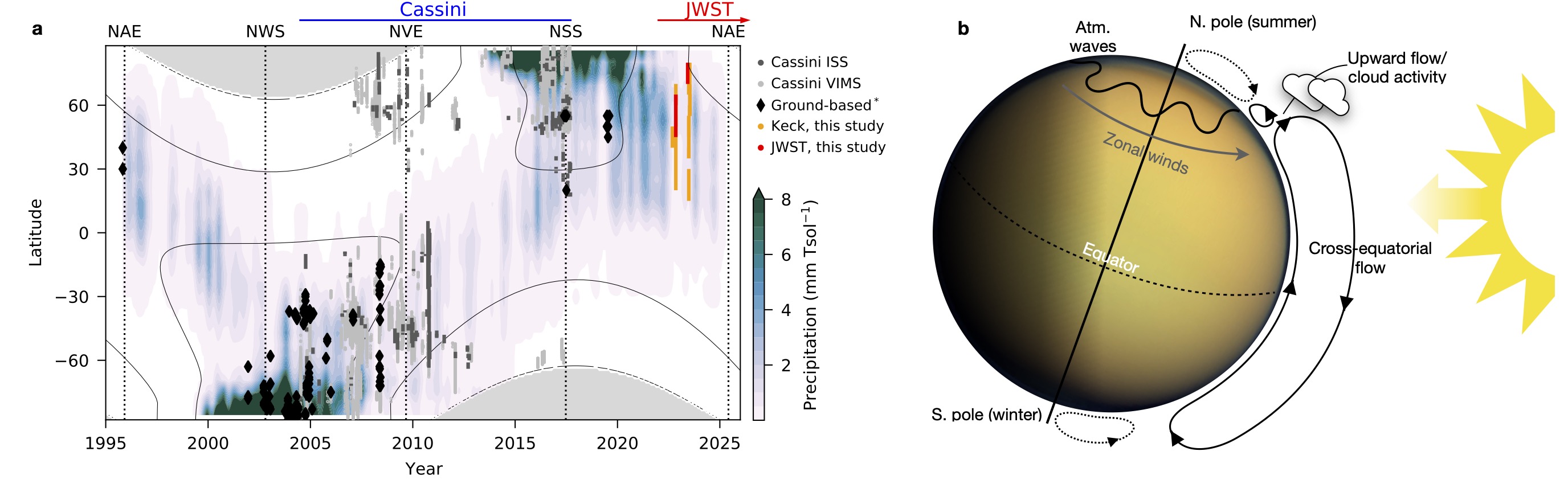} 
\caption{{\bf Seasonal distribution of methane clouds in Titan's troposphere.} (a) Latitudinal distribution of clouds observed from Cassini and ground-based observatories \citep{turtle2018titan,lemmon2019,corlies2022}, overlain on the zonal-mean precipitation distribution averaged over 40 Titan years of a simulation with TAM \citep{lora2022}. Vertical dotted lines indicate the timing of solstices (northern winter solstice and northern summer solstice, NWS and NSS) and equinoxes (northern vernal equinox and northern autumnal equinox, NVE and NAE); thin contours denote isolines of diurnally averaged top-of-atmosphere insolation, and grey shaded regions denote latitudes and times of polar night. (b) A simplified illustration of the expected atmospheric circulation during late northern summer leading to upwelling at northern mid-latitudes, which promotes cloud formation there, as observed by JWST and Keck most recently (see Figs. \ref{fig:cloud_images1} and \ref{fig:cloud_images2}).}
\label{fig:multi3}
\end{figure}


\end{document}